\documentclass[twocolumn,twocolappendix]{aastex631}

\usepackage{amsmath}
\shorttitle{Vertical Structure of PDS~70}
\shortauthors{Law et al.}
\graphicspath{{./}{figures/}}

\begin{document}

\title{Mapping the Vertical Gas Structure of the Planet-hosting PDS~70 Disk}

\author[0000-0003-1413-1776]{Charles J. Law}
\altaffiliation{NASA Hubble Fellowship Program Sagan Fellow}
\affiliation{Department of Astronomy, University of Virginia, Charlottesville, VA 22904, USA}
\affiliation{Center for Astrophysics \textbar\, Harvard \& Smithsonian, 60 Garden St., Cambridge, MA 02138, USA}

\author[0000-0002-7695-7605]{Myriam Benisty}
\affiliation{Universit\'{e} C\^{o}te d'Azur, Observatoire de la C\^{o}te d'Azur, CNRS, Laboratoire Lagrange, F-06304 Nice, France}
\affiliation{Universit\'{e} Grenoble Alpes, CNRS, IPAG, 38000 Grenoble, France}

\author[0000-0003-4689-2684]{Stefano Facchini}
\affiliation{Dipartimento di Fisica, Universit\`a degli Studi di Milano, Via Celoria 16, 20133 Milano, Italy}

\author[0000-0003-1534-5186]{Richard Teague}
\affiliation{Department of Earth, Atmospheric, and Planetary Sciences, Massachusetts Institute of Technology, Cambridge, MA 02139, USA}

\author[0000-0001-7258-770X]{Jaehan Bae}
\affiliation{Department of Astronomy, University of Florida, Gainesville, FL 32611, USA}

\author[0000-0001-8061-2207]{Andrea Isella}
\affiliation{Department of Physics and Astronomy, Rice University, 6100 Main Street, MS-108, Houston, TX 77005, USA}

\author[0000-0001-7455-5349]{Inga Kamp}
\affiliation{Kapteyn Astronomical Institute, University of Groningen, PO Box 800, 9700 AV Groningen, The Netherlands}

\author[0000-0001-8798-1347]{Karin I. \"Oberg}
\affiliation{Center for Astrophysics \textbar\, Harvard \& Smithsonian, 60 Garden St., Cambridge, MA 02138, USA}

\author[0000-0002-6278-9006]{Bayron Portilla-Revelo}
\affiliation{Kapteyn Astronomical Institute, University of Groningen, PO Box 800, 9700 AV Groningen, The Netherlands}

\author{Luna Rampinelli}
\affiliation{Dipartimento di Fisica, Universit\`a degli Studi di Milano, Via Celoria 16, 20133 Milano, Italy}



\begin{abstract}
PDS~70 hosts two massive, still-accreting planets and the inclined orientation of its protoplanetary disk presents a unique opportunity to directly probe the vertical gas structure of a planet-hosting disk. Here, we use high-spatial-resolution~(${\approx}0\farcs1$;10~au)~observations in a set of CO isotopologue lines and HCO$^+$~J=4--3 to map the full 2D~$(r,z)$~disk structure from the disk atmosphere, as traced by $^{12}$CO, to closer to the midplane, as probed by less abundant isotopologues and HCO$^+$. In the PDS~70 disk, $^{12}$CO traces a height of $z/r\approx0.3$, $^{13}$CO is found at $z/r\approx0.1$, and C$^{18}$O originates at, or near, the midplane. The HCO$^+$ surface arises from $z/r\approx0.2$ and is one of the few non-CO emission surfaces constrained with high fidelity in disks to date. In the $^{12}$CO~J=3--2 line, we resolve a vertical dip and steep rise in height at the cavity wall, making PDS~70 the first transition disk where this effect is directly seen in line emitting heights. In the outer disk, the CO emission heights of PDS~70 appear typical for its stellar mass and disk size and are not substantially altered by the two inner embedded planets. By combining CO isotopologue and HCO$^+$ lines, we derive the 2D~gas temperature structure and estimate a midplane CO snowline of ${\approx}$56-85~au. This implies that both PDS~70b and 70c are located interior to the CO snowline and are likely accreting gas with a high C/O ratio of ${\approx}$1.0, which provides context for future planetary atmospheric measurements from, e.g., JWST, and for properly modeling their formation histories.
\end{abstract}


\section{Introduction} \label{sec:intro}

Planets assemble and acquire their compositions from gas and dust in their natal protoplanetary disks, while the planet formation process is also expected to simultaneously alter the disk physical and chemical structure \citep[e.g.,][]{Cleeves15, Facchini18, Favre19}. In particular, vertical gas flows have been identified in multiple disks with suspected planets, which suggests that perturbations in the vertical gas structure of disks may be common planetary signposts \citep[e.g.,][]{Teague19Natur,Yu21,Galloway23,Izquierdo23}. However, the direct detection of planets embedded in disks remains difficult and only a handful of such systems have thus far been robustly identified \citep{Keppler18, Haffert19, Currie22, Hammond23, Wagner23}, making it difficult to conclusively assess the potential influence of giant planets on disk structure. Of these systems, PDS~70 is the only source whose protoplanetary disk is at a favorable inclination \citep[51$^{\circ}$.7;][]{Keppler19} to allow for a direct view of its vertical gas distribution and thus provides a unique opportunity to understand if and how forming planets alter disk vertical structure.

PDS~70 is a ${\sim}$5~Myr-old, K7 star \citep{Muller18} at a distance of 112~pc \citep{Gaia21} in the Upper Centaurus Lupus association \citep{Pecaut16} that hosts a gas-rich protoplanetary disk \citep{Long_Zach18, Facchini21PDS} with two, still-accreting giant planets PDS~70b and 70c that have been directly imaged at multiple wavelengths \citep{Keppler18, Muller18,  Wagner18, Christiaens19, Haffert19, Mesa19, Follette23}. PDS~70b and 70c are located at ${\approx}$22 and ${\approx}$34~au, respectively, in a near 2:1 mean motion resonance \citep{Bae19, Wang21}. Although the precise planet properties remain considerably uncertain, due in part to the presence of circumplanetary dust \citep[e.g.,][]{Isella19, Stolker20, Benisty2021}, they must be sufficiently massive (i.e., a few M$_{\rm{Jup}}$) to have carved a central gas and dust cavity seen in molecular gas, mm continuum, and NIR/scattered light \citep{Hashimoto12,Hashimoto15,Keppler19,Facchini21PDS,Portilla_Revelo22,Portilla_inprep}.

The relative proximity and inclined orientation of the PDS~70 disk makes it possible to spatially resolve emission arising from elevated regions above and below the disk midplane using high-angular-resolution observations from the Atacama Large Millimeter/submillimeter Array (ALMA) \citep[e.g.,][]{Rosenfeld13}. This then allows for the direct extraction of emission surfaces of vertically extended molecular lines using similar techniques that have now been applied to a substantial number of mid-inclination disks \citep[e.g.,][]{pinte18, Teague_19Natur, Rich21, Izquierdo21_DM1, Law21_MAPSIV, Law22, Paneque22, Paneque23, Izquierdo23, Stapper23, Law23}. Using these methods for the PDS~70 disk, we can map out the line emitting heights and conduct a detailed search for vertical perturbations, such as those driven by the embedded planets PDS~70b and 70c. 

By combining multiple optically-thick molecular lines, which trace different heights in the disk, we can also empirically derive the full two-dimensional (2D) temperature distribution \citep[e.g.,][]{Dartois03, Rosenfeld13, pinte18, Law21_MAPSIV, Leemker22, Law23}. Having a well-constrained thermal structure is crucial as it allows for the determination of important volatile snowline locations, which have a direct impact on the expected atmospheric composition of embedded planets \citep[e.g.,][]{Oberg11, Mordasini16, Molliere22}. Moreover, disk gas temperatures are vital inputs to numerical simulations and thermochemical models \citep[e.g.,][]{Bae18, Zhang21, Calahan21} and are required to infer the radial and vertical locations of molecular species, e.g., large organic molecules, that are otherwise too faint to directly map out \citep[e.g.,][]{Ilee21}. This is especially critical to determine what types of chemical reservoirs are accessible to be accreted by PDS~70b and 70c and to more generally connect planetary atmospheric composition with that of the disk environment.

Here, we extract line emission surfaces in the PDS~70 disk using high-spatial-resolution observations of CO isotopologues and HCO$^+$. In Section \ref{sec:observations_overview}, we describe the observations, imaging, and emission surface extraction process. We present the derived emission surfaces along with radial and vertical temperature profiles in Section \ref{sec:results}. In Section \ref{sec:discussion}, we explore the origins of the observed disk vertical structure and estimate the CO snowline location based on the derived thermal structure. We summarize our conclusions in Section \ref{sec:conlcusions}.

\section{Observations and analysis}
\label{sec:observations_overview}

\subsection{Observational Details}
\label{sec:archival_data}

We present long-baseline (LB) observations associated with the ALMA program (2019.1.01619.S; PI: S. Facchini) and Table \ref{tab:full_obs_program_details} lists a summary of observation dates and details. Here, we restrict our analysis to a subset of the lines presented in the chemical inventory of \citet{Facchini21PDS} based on the short-baseline (SB) data from the same program, namely, the J=2--1 line of the $^{12}$CO, $^{13}$CO, and C$^{18}$O isotopologues, which are sufficiently bright and vertically extended to allow for extraction of their vertical structure. Full details about these SB data, including the spectral set-up, are presented in \citet{Facchini21PDS} and the LB data for the remaining molecules will be published in forthcoming papers. We also make use of existing archival observations of the $^{12}$CO J=3--2 and HCO$^+$ J=4--3 lines in the PDS~70 disk compiled from ALMA programs 2015.1.00888.S (PI: E. Akiyama) and 2017.A.00006.S (PI: M. Keppler), which were presented in \citet{Long_Zach18, Keppler19, Facchini21PDS}. We also include unpublished LB data of the same lines from 2018.A.00030.S (PI: M. Benisty). See \citet{Benisty2021} for additional details about each of these programs. We include HCO$^+$ in our analysis, due to its vertically-elevated emission and the availability of existing high-angular-resolution line data. Since HCO$^+$ typically shows bright, optically-thick emission in disks \citep[e.g.,][]{Booth19, Huang20}, it is thus also useful for mapping the thermal structure of PDS~70 in combination with CO isotopologues.

\begin{deluxetable*}{ccccccccccc}[h]
\tablecaption{Details of ALMA LB Observations of PDS~70 \label{tab:full_obs_program_details}}
\tablewidth{0pt}
\tablehead{
\colhead{UT Date} & \colhead{Ants.} & \colhead{Int.} & \colhead{Baselines} & \colhead{Res.} & \colhead{M.R.S.} & \colhead{PWV} & \multicolumn2c{Calibrators} \\ \cline{8-9} 
\colhead{} & \colhead{} & \colhead{(min)} & \colhead{(m)} & \colhead{($^{\prime \prime}$)} & \colhead{($^{\prime \prime}$)}  & \colhead{(mm)} & \colhead{Flux/Bandpass} & \colhead{Phase} 
}
\startdata
2021-07-14 & 44 & 75.9 & 15.0-3396.4 & 0.1 & 1.8 & 1.7 & J1427-4206 & J1352-4412 \\
2021-07-15 & 29 & 75.6 & 15.0-3396.4 & 0.1 & 1.8 & 1.4 & J1427-4206 & J1352-4412  \\
2021-07-15 & 43 & 74.6 & 15.0-3396.4 & 0.1 & 1.8 & 1.1 & J1427-4206 & J1352-4412 \\
2021-07-16 & 46 & 76.4 & 15.0-3638.2 & 0.1 & 2.0 & 1.2 & J1427-4206 & J1352-4412 
\enddata
\end{deluxetable*}

\begin{figure*}[!htpb]
\centering 
\includegraphics[width=\linewidth]{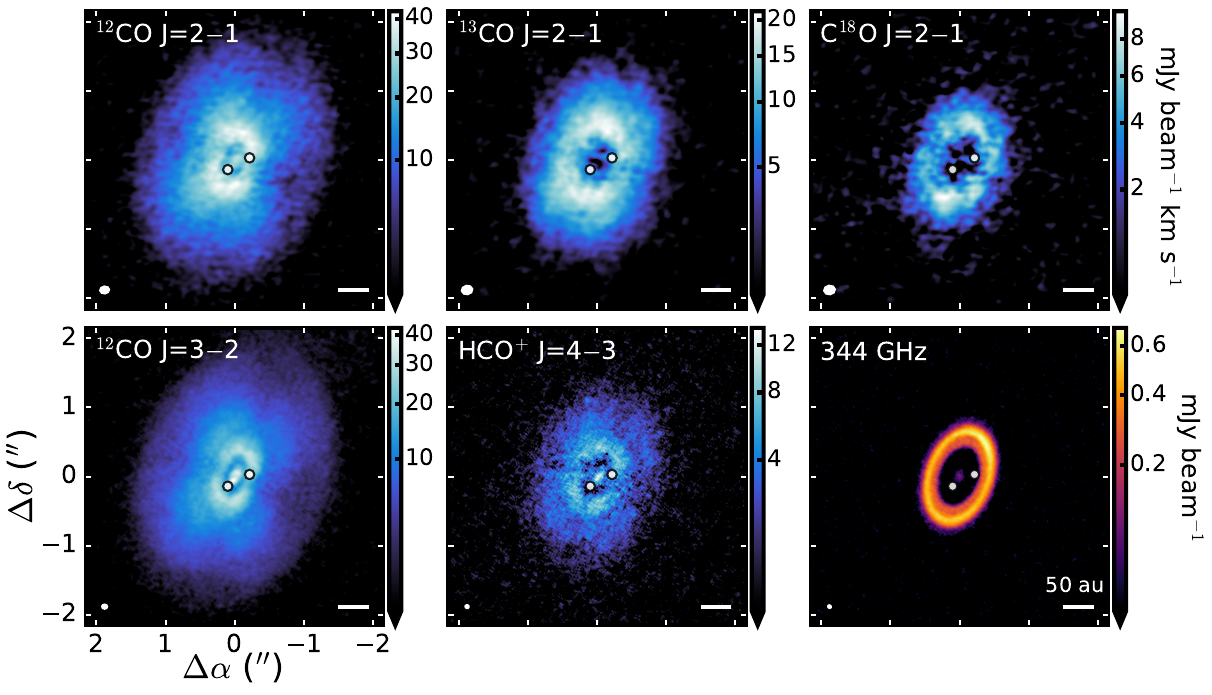} 
\caption{CO isotopologue and HCO$^+$ zeroth moment maps and continuum images (from top left to bottom right) of the PDS~70 disk. Panels for each disk have the same field of view. The locations of the planets PDS~70b and 70c are marked in each panel \citep{Wang21}. Color stretches were individually optimized and applied to each panel to increase the visibility of outer disk structure. The synthesized beam and a scale bar indicating 50~au is shown in the lower left and right corner, respectively, of each panel. The 344~GHz continuum is from \citet{Isella19}. Table \ref{tab:image_info} lists details about each image.}
\label{fig:figure1}
\end{figure*}

\subsection{Self Calibration and Imaging}
\label{sec:selfcal_imaging}

The three J=2--1 CO isotopologue lines are included in one spectral setup of the \citet{Facchini21PDS} program. For these Band 6 data, we followed the same self-calibration procedure, in line with \citet{Andrews18,Benisty2021}. The self-calibration was performed with CASA \texttt{v5.8} \citep{McMullin_etal_2007, CASA}. We first re-aligned the data of the individual Execution Blocks (EBs) to a common disk center with the \texttt{fixvis} and \texttt{fixplanets} tasks. The center was computed by fitting an ellipse to the outer ring \citep[also clearly discernible in the SB data, see][]{Facchini21PDS}. This approach had previously performed well for the self-calibration of the Band 7 data of the PDS~70 disk by \citet{Benisty2021}. After flagging the lines listed in \citet{Facchini21PDS} within $\pm15\,$km\,s$^{-1}$ from the line center, we spectrally averaged the data into 250\,MHz channels to create pseudo-continuum ms tables. Individual EBs were rescaled in amplitude to provide the same disk flux density, after verifying that phase de-coherence was not artificially reducing it. The SB data were self-calibrated first, and then concatenated to the LB data before running a new self-calibration. The \texttt{tclean} model was used as the source model for each step, where \texttt{tclean} was performed over an elliptical mask with a 1\farcs5 radius and inclination and position angle as in \citet{Facchini21PDS}. Four rounds of phase-only and one round of amplitude self-calibration were performed on the SB data, and three rounds of phase-only and one of amplitude on the LB data, respectively. Before amplitude calibration, we \texttt{tcleaned} the data with a deep $1\sigma$ threshold. Using Briggs weighting with \texttt{robust}=0.5, the synthesized beam was $0\farcs15\times0\farcs12$ (PA=$-82.9^{\circ}$). The resulting 233~GHz continuum image had a flux density of 58.3\,mJy within the CLEAN mask \citep[in agreement with][]{Facchini21PDS}, with an rms of $12.5\,\mu$Jy\,beam$^{-1}$. We finally applied the gain solutions to the full spectral data.

We obtained archival Band 7 observations of the $^{12}$CO J=3--2 and HCO$^+$ J=4--3 lines from the datasets presented in \citet{Benisty2021}, for which we applied the same gain solutions to the line spectral windows. For both the Band 6 and Band 7 lines, we continuum-subtracted the data with the \texttt{contsub} task using a first-order polynomial.

We imaged the five lines discussed in this paper with CASA \texttt{v6.2}. The channel maps were restored using \texttt{tclean} with a threshold of $3.5\sigma$. Table~\ref{tab:image_info} lists the imaging parameters of all cubes. We used Keplerian masks for the channels generated from the \texttt{keplerian\_mask} code \citep{rich_teague_2020_4321137}, where we assumed the disk to be extended $3\arcsec$ in radius, and we conservatively assumed that all emission is as elevated as $^{12}$CO. Given the focus of the paper, extracting accurate brightness temperatures is imperative. We thus applied the so-called JvM correction to the final images \citep{JvM95,Czekala21} to more accurately recover the low surface brightness intensities in the disk outer regions, and of weaker lines (e.g., C$^{18}$O J=2--1). The effective rms, which can be underestimated when applying the JvM correction \citep{Casassus22}, is not relevant for the analysis of this paper. The full image cube channel maps are availabe in Appendix \ref{sec:appendix_channel_maps}. We also imaged the non-continuum-subtracted line data by adopting the same imaging parameters as the line only emission image cubes. The non-continuum-subtracted image cubes are required for the calculation of gas temperatures (Section \ref{sec:gas_temperatures}). Overall, the ability to extract line emitting heights depends on having sufficiently high angular resolution, spectral resolution, and line sensitivities. Considering this, we selected the imaging parameters that best suited surface extraction for this work. Table \ref{tab:image_info} lists the properties of the image cubes selected for analysis of the vertical line emission structure. 

Figure \ref{fig:figure1} shows an overview the CO isotopologue and HCO$^+$ line emission velocity-integrated intensity, or ``zeroth moment," maps, and the 344~GHz continuum emission. The continuum image is taken from \citet{Isella19}. We generated zeroth moment maps of line emission from the image cubes using \texttt{bettermoments} \citep{Teague18_bettermoments} with the same Keplerian masks employed during CLEANing and with no flux threshold for pixel inclusion to ensure accurate flux recovery.

\subsection{Emission Surface Extraction}\label{sec:methods_sub_surfextr}

We derived vertical emission heights from the line emission image cubes using the \texttt{disksurf} \citep{disksurf_Teague} Python code, which is based on the methodology originally presented in \citet{pinte18}. We closely followed the methods outlined in \citet{Law21_MAPSIV}, which we briefly summarize below.

\begin{deluxetable*}{llccccccccc}[h]
\tabletypesize{\footnotesize}
\tablecaption{Image Cube Properties\label{tab:image_info}}
\tablehead{
\colhead{Transition}  & \colhead{Beam} & \colhead{JvM $\epsilon$\tablenotemark{a}} & \colhead{\texttt{robust}} & \colhead{Chan. $\delta$v} & \colhead{RMS} & \colhead{R$_{\rm{size}}$\tablenotemark{b}} & \colhead{ALMA}  \vspace{-0.25cm}\\
\colhead{} & \colhead{ ($^{\prime \prime} \times ^{\prime \prime}$, $\deg$)} & & & \colhead{(km~s$^{-1}$)} &  \colhead{(mJy~beam$^{-1}$)} & \colhead{(au)} & \colhead{Project Code(s)} }
\startdata
$^{12}$CO J=2--1 & 0.13 $\times$ 0.10, $-$83.7  & 0.61 & 0.25 & 0.30 & 0.93 & 209~$\pm$~4 & P2019 \\
$^{13}$CO J=2--1 & 0.15 $\times$ 0.12, $-$80.0 & 0.46 & 0.5 & 0.20 & 0.62 & 167~$\pm$~5 & P2019   \\
C$^{18}$O J=2--1 & 0.16 $\times$ 0.13, $-$82.4 & 0.46 & 0.5 & 0.40 & 0.42 & 142~$\pm$~7 & P2019 \\
$^{12}$CO J=3--2 & 0.07 $\times$ 0.06, 87.1  & \ldots & 1.0 & 0.43 & 0.26 & 210~$\pm$~2 & P2015, P2017, P2018 \\
HCO$^+$ J=4--3   & 0.05 $\times$ 0.04, 69.3 & \ldots & 1.0 & 0.43 & 0.31  & 165~$\pm$~3 & P2015, P2018 \\
\enddata
\tablenotetext{a}{The ratio of the CLEAN beam and dirty beam effective area used to scale image residuals to account for the effects of non-Gaussian beams. See Section \ref{sec:selfcal_imaging} and \citet{JvM95, Czekala21} for further details.}
\tablenotetext{b}{Disk size (R$_{\rm{size}}$) was computed as the radius which encloses 90\% of the total disk flux.}
\end{deluxetable*}

\begin{figure*}[!htpb]
\centering
\includegraphics[width=1\linewidth]{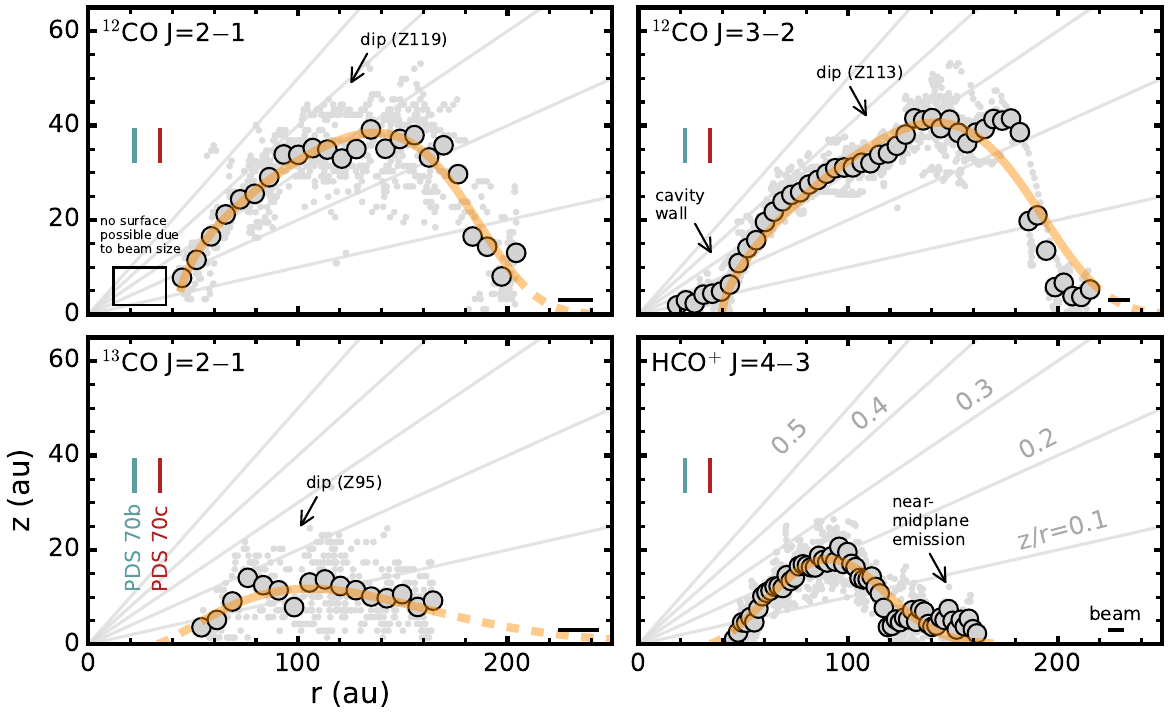}
\caption{CO isotopologue and HCO$^+$ emission surfaces for the PDS~70 disk. Large gray points show radially-binned surfaces and small, light gray points represent individual measurements. The orange lines show the exponentially-tapered power law fits from Table \ref{tab:emission_surf}. The solid lines show the radial range used in the fitting, while the dashed lines are extrapolations. Lines of constant $z/r$ from 0.1 to 0.5 are shown in gray. Vertical substructures and the locations of PDS~70b and 70c are marked in each panel \citep{Wang21}. The lack of $^{12}$CO J=2--1 heights in the inner 50~au (square box) is due to insufficient angular resolution rather than the absence of emission at these radii. The FWHM of the major axis of the synthesized beam is shown in the bottom right corner of each panel. The emission surfaces shown in this figure are available as Data behind the Figure.}
\label{fig:figure_gallery_r_v_z_12CO}
\end{figure*}

We first adopted a PA=160$^{\circ}$.4 and inc=51$^{\circ}$.7 for the PDS~70 disk \citep{Keppler19, Facchini21PDS} and then extracted a deprojected radius $r$, emission height $z$, surface brightness $I_{\nu}$, and channel velocity $v$ for each pixel associated with the emitting surface. We then employed an iterative fitting approach, which further refines the extracted surfaces using between one to five subsequent iterations depending on the line. To ensure the robustness of the extracted points, we also filtered those pixels with unphysical $z$/$r$ values based on the expected disk structure and removed those points with low surface brightnesses (2-3$\times$rms). At each step, we visually confirmed the fidelity of the derived emission surfaces. To further reduce scatter in the extracted surfaces, we also performed each of the same two types of binning (radially binned and moving averages) as in \citet{Law21_MAPSIV} with half-beam spacings. 

We then fitted parametric models to all line emission surfaces using an exponentially-tapered power law, which captures the inner flared surfaces and the plateau and turnover regions at larger radii \citep[e.g.,][]{Teague_19Natur}. We adopted the following functional form:

\begin{equation} \label{eqn:exp_taper}
z(r) = z_0 \times \left( \frac{r - r_{\rm{cavity}}}{1^{\prime \prime}} \right)^{\phi} \times \exp \left(- \left[ \frac{r}{r_{\rm{taper}}} \right]^{\psi} \right)
\end{equation}

\noindent This is a similar form employed by \citet{Law21_MAPSIV} but here we include an additional term r$_{\rm{cavity}}$ to describe the inner gas cavity of the PDS~70 disk. 

We used the Monte Carlo Markov Chain (MCMC) sampler implemented in \texttt{emcee} \citep{Foreman_Mackey13} to estimate the posterior probability distributions for: $z_0$, $\phi$, r$_{\rm{taper}}$, $r_{\rm{cavity}}$, and $\psi$. Each ensemble used 64 walkers with 1000 burn-in steps and an additional 500 steps to sample the posterior distribution function. The posteriors were approximately Gaussian with no significant degeneracies between parameters. Table \ref{tab:emission_surf} shows the median values of the posterior distribution, with uncertainties given as the 16th and 84th percentiles, as well as the radial range, r$_{\rm{fit,\,max}}$, considered for each surface.

\begin{deluxetable*}{lccccccc}
\tablecaption{Parameters for CO Isotopologue and HCO$^+$ Emission Surface Fits\label{tab:emission_surf}}
\tablewidth{0pt}
\tablehead{
 \colhead{Line} & &  \multicolumn5c{Exponentially-Tapered Power Law} & \colhead{Char. z/r\tablenotemark{a}} \\ \cline{3-7}
 \colhead{} & \colhead{r$_{\rm{fit,\,max}}$ ($^{\prime \prime}$)} & \colhead{$z_0$ ($^{\prime \prime}$)} & \colhead{$\phi$} & \colhead{r$_{\rm{cavity}}$ ($^{\prime \prime}$)} & \colhead{$r_{\rm{taper}}$ ($^{\prime \prime}$)} &\colhead{$\psi$}}
\startdata
$^{12}$CO J=2$-$1 & 1.80 & 0.41$^{+0.02}_{-0.02}$ & 0.51$^{+0.08}_{-0.06}$ & 0.37$^{+0.02}_{-0.03}$ & 1.27$^{+0.03}_{-0.02}$ & 5.74$^{+0.97}_{-0.84}$ & 0.32$^{+0.01}_{-0.06}$\\
$^{13}$CO J=2$-$1 & 1.45 & 0.41$^{+1.30}_{-0.24}$ & 1.29$^{+1.16}_{-0.87}$ & 0.28$^{+0.21}_{-0.21}$ & 0.74$^{+0.52}_{-0.37}$ & 1.61$^{+1.28}_{-0.75}$ & 0.09$^{+0.04}_{-0.02}$\\
$^{12}$CO J=3$-$2 & 1.90 & 0.43$^{+0.03}_{-0.05}$ & 0.63$^{+0.05}_{-0.13}$ & 0.35$^{+0.00}_{-0.02}$ & 1.34$^{+0.02}_{-0.03}$ & 5.40$^{+4.09}_{-1.80}$ & 0.32$^{+0.02}_{-0.04}$\\
HCO$^+$ J=4$-$3 & 1.55 & 0.70$^{+0.20}_{-0.23}$ & 1.48$^{+0.89}_{-0.57}$ & 0.29$^{+0.11}_{-0.18}$ & 0.65$^{+0.14}_{-0.10}$ & 3.00$^{+0.91}_{-0.70}$ & 0.18$^{+0.02}_{-0.09}$\\
\enddata
\tablenotetext{a}{Characteristic $z/r$ is computed as the mean and 16th to 84th percentile range within 80\% of the r$_{\rm{cutoff}}$ as in \citet{Law22,Law23}.}
\end{deluxetable*}

\begin{figure*}
\centering
\includegraphics[width=0.75\linewidth]{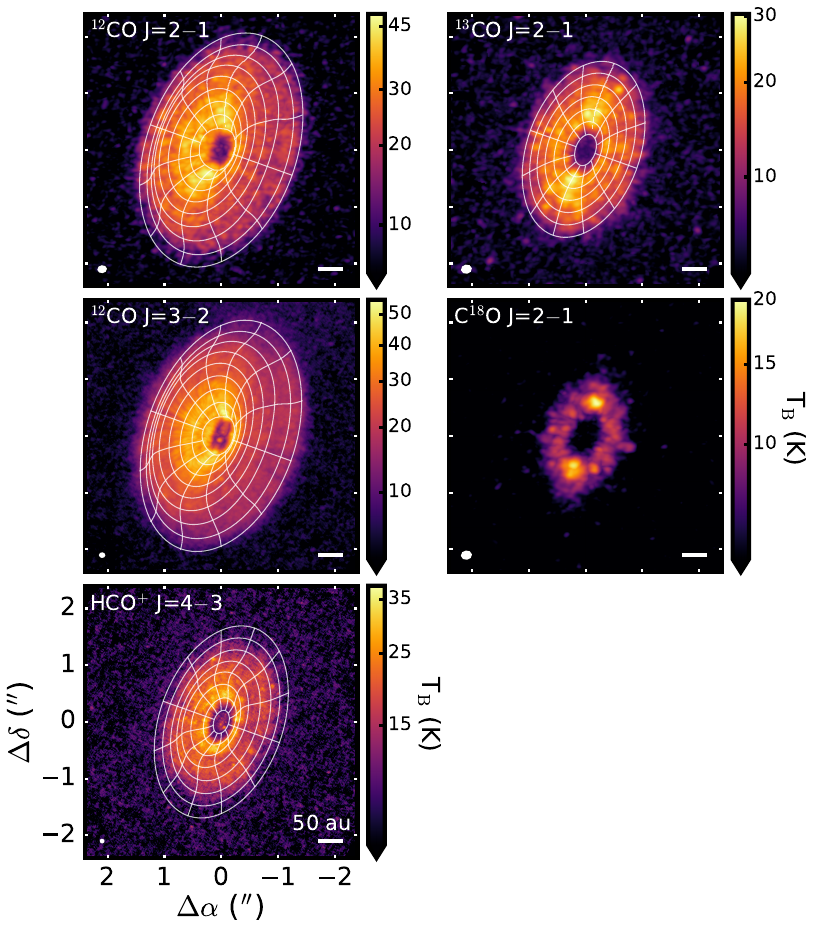}
\caption{Peak intensity maps with overlaid contours showing the fitted emission surfaces, as listed in Table \ref{tab:emission_surf}. No surface could be derived for the C$^{18}$O J=2--1 line, which is consistent with emission originating from near the midplane. The synthesized beam and a scale bar indicating 50~au is shown in the lower left and right corner, respectively, of each panel.}
\label{fig:overlaid_Fnu}
\end{figure*}

\begin{deluxetable*}{llccccccc}
\tablecaption{Properties of Vertical Substructures \label{tab:vertical_substructures}}
\tablewidth{0pt}
\tablehead{\colhead{Line} & \colhead{Feature} & \colhead{$r_0$ (au)} & \colhead{Width (au)} & \colhead{$\Delta$z (au)} & \colhead{Depth}}
\startdata
$^{12}$CO J=2$-$1 & Z119 & 119 $\pm$ 1 & 14 $\pm$ 7 &  2 $\pm$ 1 & 0.07 $\pm$ 0.06\\
$^{12}$CO J=3$-$2 & Z42 & 42 $\pm$ 0.1 & 21 $\pm$ 4 &  6 $\pm$ 0.2 & 0.48 $\pm$ 0.07\\
 & Z113 & 113 $\pm$ 1 & 25 $\pm$ 6 &  3 $\pm$ 1 & 0.10 $\pm$ 0.08\\
$^{13}$CO J=2$-$1 & Z95 & 95 $\pm$ 2 & 20 $\pm$ 0.3 &  4 $\pm$ 1 & 0.28 $\pm$ 0.14\\
\enddata
\end{deluxetable*}

\section{Results} \label{sec:results}

\subsection{Emission Surfaces in the PDS~70 Disk} \label{sec:overview_emission_surfaces}

Figure \ref{fig:figure_gallery_r_v_z_12CO} shows the derived emission surfaces in the PDS~70 disk. All of the CO isotopologue lines show characteristic emission surface profiles, i.e., a sharply-rising inner component, followed by subsequent plateau and turnover phases toward larger radii \citep[e.g.,][]{Teague_19Natur, Law21_MAPSIV}. Due to the high sensitivity and angular resolution of the observations combined with the iterative surface fitting procedure of \texttt{disksurf}, we are also able to confidently track emission heights even at large disk radii.

\begin{figure*}[!t]
\centering
\includegraphics[width=\linewidth]{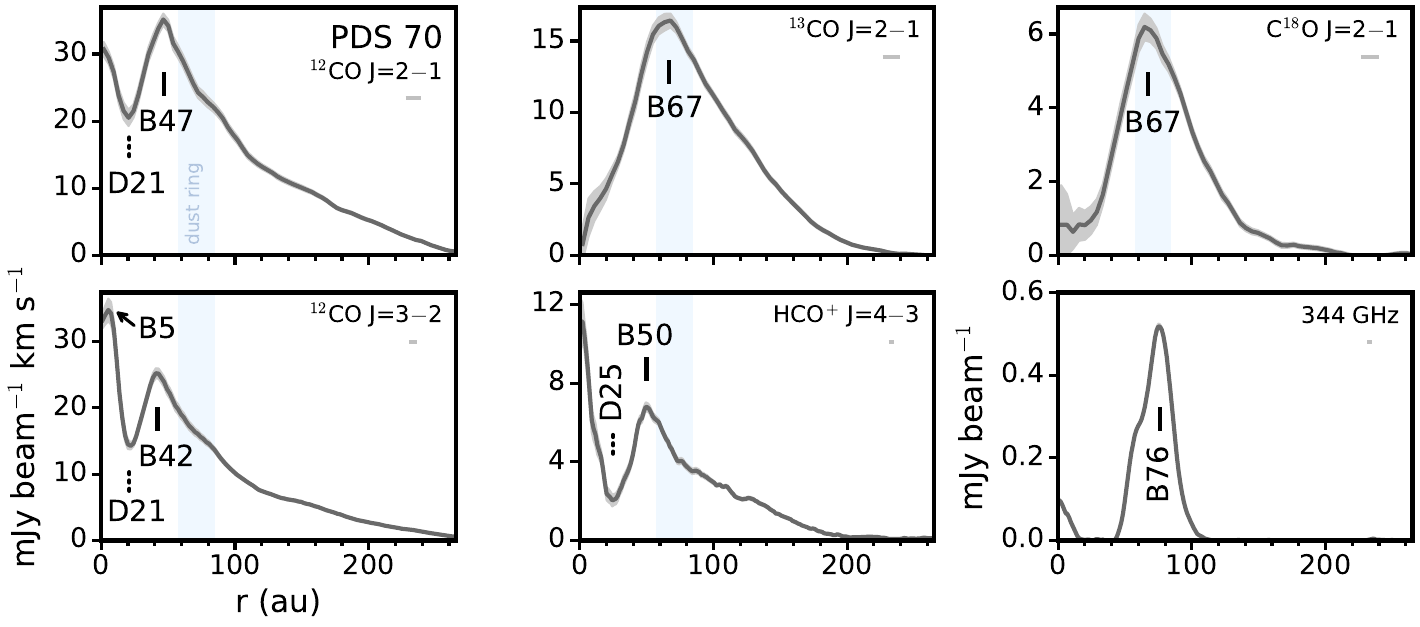}
\caption{Deprojected radial integrated intensity profiles lines and the 344~GHz continuum in the PDS~70 disk. Shaded regions show the 1$\sigma$ uncertainty. Solid lines mark emission rings and dotted lines indicate gaps. The FWHM of the synthesized beam is shown by a horizontal bar in the upper right corner of each panel.}
\label{fig:radial_profiles}
\end{figure*}

Both the J=3--2 and J=2--1 lines of $^{12}$CO show altitudes of $z/r \approx 0.3$ and reach a maximum height of approximately 40~au at a radius of 150~au. The $^{12}$CO J=3--2 heights are generally consistent with those derived in \citet{Keppler19}, who fit a single power-law surface profile to the Keplerian rotation map but did not extract individual heights. The $^{12}$CO J=3--2 emission surface shows evidence for vertical variation due to the cavity wall at ${\approx}$45~au. From a radius of ${\approx}$15-40~au, the surface is quite flat ($z/r\approx0.1$) and located near the midplane, but then abruptly increases height at a radius of ${\approx}$45~au up to $z/r\approx0.3$. We are unable to extract emission heights for the $^{12}$CO J=2--1 line in the inner 50~au (square box in Figure \ref{fig:figure_gallery_r_v_z_12CO}) despite $^{12}$CO emission being clearly detected in the inner disk (Figure \ref{fig:figure1}). This is likely due to the larger beam size of the J=2--1 observations (3 times larger in beam area) rather than an intrinsic difference in emitting heights, especially since the J=2--1 and J=3--2 $^{12}$CO lines have typically been observed to originate from the same heights in other disks \citep{Law23}. The $^{13}$CO J=2--1 line originates from a deeper layer in the disk ($z/r\approx0.1$). We do not resolve the vertical structure of the C$^{18}$O emission, which suggests that C$^{18}$O arises at, or near, the disk midplane. In addition to CO isotopologues, we also measured the HCO$^+$ J=4--3 emitting surface, which lies at $z/r\approx0.2$. Beyond ${\approx}$100~au, the HCO$^+$ surface shows an extended plateau of flat, near-midplane emission ($z/r \lesssim 0.1$), which is not observed in any other line. At ${<}$100~au, the HCO$^+$ arises from higher altitudes than $^{13}$CO, but at ${>}$100au, HCO$^+$ is coming from lower altitudes than $^{13}$CO.

In addition to the cavity wall seen in $^{12}$CO J=3--2, each of the $^{12}$CO and $^{13}$CO surfaces shows a shallow vertical dip between ${\approx}$90-120~au. We confirmed that each of substructures are also detected in emission surfaces extracted using the non-continuum-subtracted image cubes and with no filtering of low surface brightness points, and thus, are not artifacts of the continuum subtraction or surface extraction process. To quantify the significance of these features, including the cavity wall, we follow the fitting procedure outlined in \citet{Law21_MAPSIV}. In brief, we remove a local baseline around each vertical substructure and then fit for its radial position, width, and depth. Table \ref{tab:vertical_substructures} shows the properties of each substructure, which are labeled with a “Z” to indicate that they are vertical variations followed by their radial location rounded to the nearest whole number in astronomical units. The cavity wall at Z42 is the deepest vertical feature, followed by the Z95 dip in $^{13}$CO, while the dips at Z119 and Z113 in the $^{12}$CO surfaces are the most shallow.

In Figure \ref{fig:overlaid_Fnu}, we overlay the inferred emission surfaces on peak line intensity maps to better illustrate their 3D geometries. We generated all peak intensity maps with the `quadratic' method of \texttt{bettermoments} using the full Planck function.

\subsection{Radial Disk Chemical Substructure} \label{sec:vertical_substr_vs_mm_cont}

While several molecular lines considered here have been previously observed toward the PDS~70 disk \citep[e.g.,][]{Long_Zach18, Facchini21PDS}, these new data represent improvements in both angular resolution and sensitivity of more than a factor of two. Thus, in addition to mapping the vertical gas structure, we can also constrain the presence of small-scale radial substructures at the highest spatial resolution to date. To do so, we computed azimuthally-averaged radial profiles using the \texttt{radial\_profile} function in the \texttt{GoFish} Python package \citep{Teague19JOSS} to deproject the zeroth moment maps. During deprojection, we incorporated the derived emission surfaces listed in Table \ref{tab:emission_surf}. This is particularly important for highly elevated surfaces, e.g., $^{12}$CO, to derive accurate radial locations of line emission substructures \citep[e.g.,][]{Rosotti21, Law21_MAPSIII}. Figure \ref{fig:radial_profiles} shows the resultant radial profiles.

The location of line emission substructures are labeled according to their radial location rounded to the nearest au following standard nomenclature \citep[e.g.,][]{Huang18, Law21_MAPSIII}. The $^{12}$CO lines show centrally-peaked profiles with a dip at 21~au, followed by an emission peak between 40-50~au and an extended plateau out to ${\approx}$250~au. The higher resolution $^{12}$CO J=3--2 profile resolves an inner ring at ${\approx}$5~au, indicating the possible presence of small-scale inner disk substructure. The $^{13}$CO and C$^{18}$O J=2--1 lines instead show a central gap and broad ring at 67~au. The HCO$^+$ profile is more similar to that of the $^{12}$CO lines with a dip at 25~au and ring at 50~au. However,  HCO$^+$ is much more sharply centrally peaked with no indications of a central dip in the inner few au, despite the HCO$^+$ having the highest spatial resolution. Overall, the radial morphologies are consistent with existing lower resolution observations \citep{Long_Zach18, Facchini21PDS}, namely that $^{12}$CO and HCO$^+$ peak interior to the mm dust ring and $^{13}$CO and C$^{18}$O are co-located with the mm continuum peak. As discussed in \citet{Facchini21PDS}, these differences are likely due to varying line optical depths and the presence of high temperatures and illumination at the edge of the cavity wall.

We also computed the disk size of each emission line by determining the radius in which 90\% of the total flux is contained \citep[e.g.,][]{Tripathi17, Ansdell18}. Table \ref{tab:image_info} lists the derived values. The $^{12}$CO lines have have sizes of ${\approx}$200~au, while $^{13}$CO and HCO$^+$ have sizes of ${\approx}$165~au. The C$^{18}$O is the most compact line at ${\approx}$140~au. The mm continuum edge of the PDS~70 disk is approximately 100~au. This implies gas-to-dust sizes of ${\approx}$1.4-2, which are typical, albeit on the lower end, for large, resolved disks \citep[e.g.,][]{Law21_MAPSIII,Sanchis21,Long22}.

\subsection{Comparison with NIR Scattering Surface} \label{sec:comparison_NIR_rings}

The PDS~70 disk has been extensively observed at NIR wavelengths, which revealed an elliptical ring, inner disk component, and asymmetrical feature to the northwest of the star due to either a double ring or planet-disk interactions \citep[e.g.,][]{Hashimoto12, Keppler18, Muller18, Mesa19, Juillard22}. Here, we only focus on the scattering surface in the outer disk, which is tracing the small dust grain population, and compare this to the derived line emission surfaces. Only a handful of disks have independent measurements of both NIR and gas heights, and as a result, the general relationship between the vertical distribution of gas and small dust in disks is not yet well established.

\begin{figure}[h]
\centering
\includegraphics[width=\linewidth]{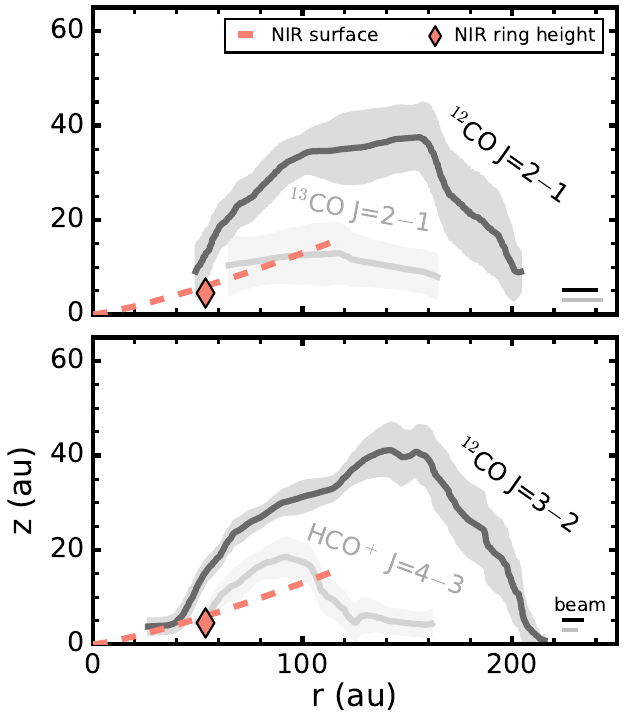}
\caption{Emission surfaces of $^{12}$CO and $^{13}$CO J=2--1 (top) and $^{12}$CO J=3--2 and HCO$^+$ J=4--3 (bottom) versus the NIR scattering surface from \citet{Keppler18} in the PDS~70 disk. The red marker shows the individual height measurement of the PDS~70 NIR ring derived in this work. The lines show the moving average surfaces and gray shaded regions show the 1$\sigma$ uncertainty. The FWHM of the major axis of the synthesized beam is shown in the bottom right corner of each panel.}
\label{fig:NIR_surface}
\end{figure}

Figure \ref{fig:NIR_surface} shows the power-law NIR surface derived in \citet{Keppler18} versus the extracted molecular gas heights. We also directly measured the height of the scattered light ring at ${\approx}$54~au in the PDS~70 disk by fitting an ellipse to the peak flux using the 2016 SPHERE IRDIS J-band, Q$_{\phi}$ polarimetric image from \citet{Keppler18}. To do so, we followed the same approach outlined in \citet{Rich21} and described in detail in Appendix C of \citet{Law23}. We derived a height of $4.5 \pm 0.1$~au at a radius of $53.7 \pm 0.1$~au, which agrees well with the overall power-law trend from \citet{Keppler18} and is shown as a diamond marker in Figure \ref{fig:NIR_surface}. The NIR surface is located at the same vertical height as the $^{13}$CO surface over the entire radial range of the NIR disk. Existing measurements in other disks show that the NIR surface typically lies between the $^{12}$CO and $^{13}$CO layers, but with considerable diversity in the relative heights of the small dust grains \citep[e.g.,][]{Rich21, Law22, Law23, Paneque23}. In particular, the well-studied IM~Lup and HD~163296 disks show comparable heights in $^{12}$CO gas and small dust grains within 100~au, but significantly more elevated $^{12}$CO emitting heights at larger radii \citep{Rich21, Paneque23}. Given that NIR emission in PDS~70 only extends to ${\approx}$115~au \citep{Keppler18}, it is thus possible that the PDS~70 disk shows slightly lower than expected small dust heights compared to that of $^{12}$CO. However, we caution that only a small number of disks have had their CO isotopologue and small dust heights jointly mapped and additional disk observations are required to better place PDS~70 in context.

\begin{figure*}[p]
\centering
\includegraphics[width=\linewidth]{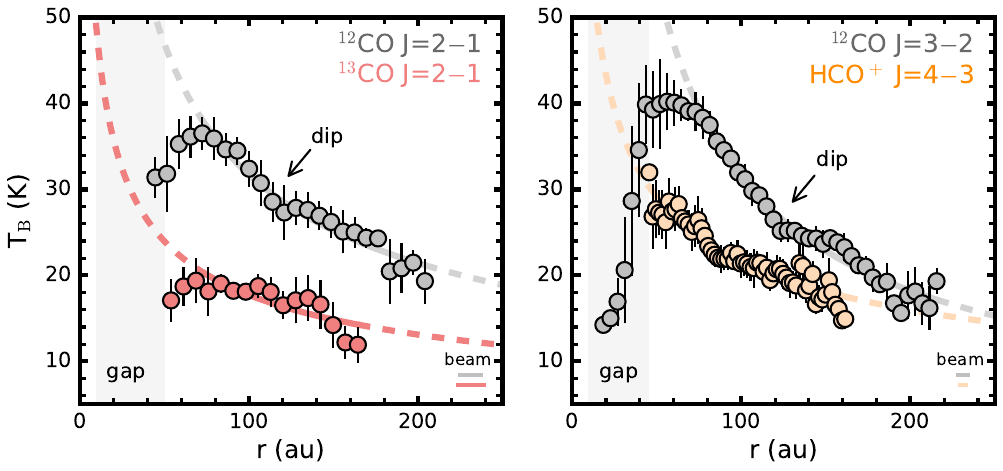}
\caption{CO, $^{13}$CO, and HCO$^+$ radial brightness temperature profiles of the PDS~70 disk. These profiles represent the mean temperatures computed by radially binning the individual measurements, similar to the procedure used to compute the radially-binned surfaces (see Section \ref{sec:gas_temperatures}). Vertical lines show the 1$\sigma$ uncertainty, given as the standard deviation of the individual measurements in each bin. The solid gray lines show the power fits from Table \ref{tab:radial_temperature_plaw_fits}, while the dashed lines are extrapolations. The FWHM of the major axis of the synthesized beam is shown in the bottom right corner of each panel.}
\label{fig:figure_temp}
\end{figure*}

\begin{figure*}[p]
\centering
\includegraphics[width=\linewidth]{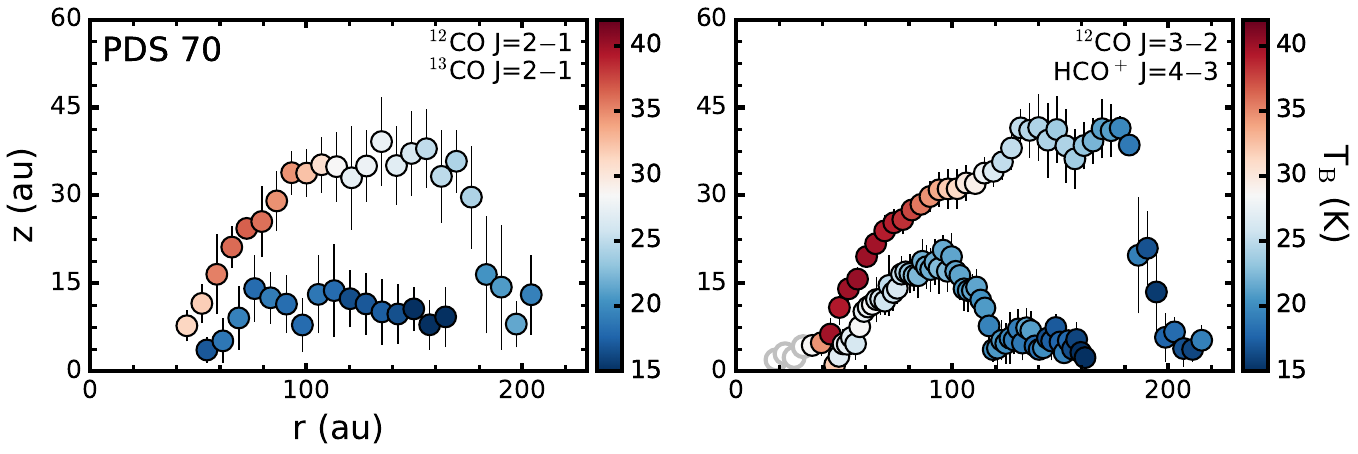}
\caption{2D temperature profiles of the $^{12}$CO and $^{13}$CO J=2--1 (left) and $^{12}$CO J=3--2 and HCO$^+$ J=4--3 (right) lines in the PDS~70 disk. Points are those from the binned surfaces and error bars are the 1$\sigma$ uncertainties in $z$. Temperature measurements with radii near the emission gap (Figure \ref{fig:figure_temp}) are marked by hollow markers and not used in the 2D temperature fits. The 2D temperature profiles shown in this figure are available as Data behind the Figure.}
\label{fig:2D_temp_surfaces}
\end{figure*}

\subsection{Disk Thermal Structure} \label{sec:gas_temperatures}

We extracted radial (Section \ref{sec:radial_temperatures}) and vertical (Section \ref{sec:2D_temps}) gas temperatures for the PDS~70 disk using $^{12}$CO, $^{13}$CO, and HCO$^+$. To do so, we followed the procedures of \citet{Law21_MAPSIV}. We assume that all lines are optically thick and in LTE, which is expected to be the case at the typical densities and temperatures of protoplanetary disks \citep[e.g.,][]{Weaver18}. We also assume that the line emission fills the beam due to the high angular resolution of the observations. We then repeated the surface extractions as in Section \ref{sec:methods_sub_surfextr} on the non-continuum-subtracted image cubes, which ensures that we do not underestimate the line intensity along lines of sight containing strong continuum emission \citep[e.g.,][]{Boehler17}. We then converted the peak surface brightness I$_{\nu}$ of each extracted pixel to a brightness temperature using the full Planck function and assume the resulting brightness temperature is equal to the local gas temperature. 

All subsequent radial and 2D gas temperature distributions are taken directly from individual surface measurements, rather than from mapping peak brightness temperatures back onto derived emission surfaces (i.e., Figure \ref{fig:overlaid_Fnu}) or radially-deprojecting peak intensity maps (see Appendix \ref{sec:appendix_Fnu_profiles}). Hence, we only consider the brightness temperature of those pixels with derived emission heights.

\subsubsection{Radial Temperature Profiles} \label{sec:radial_temperatures}

Figure \ref{fig:figure_temp} shows the radial temperature distributions along the emission surfaces. The $^{12}$CO lines trace the warmest temperatures, followed by HCO$^+$ and then $^{13}$CO. For all lines, the temperatures generally increase toward the central star, except in the inner ${\approx}$60-70~au where the temperatures of the CO lines plateau or decrease. This is unsurprising since the CO lines all either show a central cavity or deep gap in the inner disk, which means that at these radii, the lines are likely no longer optically thick and thus do not trace the true gas temperature. We also confirmed that the radial temperatures extracted directly from the emission surfaces in Figure \ref{fig:figure_temp} show excellent agreement with the peak intensity profiles generated from Figure \ref{fig:overlaid_Fnu}, which are shown in Appendix \ref{sec:appendix_Fnu_profiles}.

The J=2--1 and J=3--2 lines of $^{12}$CO are generally comparable in temperature, but the J=3--2 line has a peak temperature of ${\gtrsim}$40~K that is a few K warmer than that of the J=2--1 line, which is likely due to beam dilution at smaller radii, since the J=2--1 beam is more than twice as large than that of the J=3--2 line. While the $^{13}$CO and HCO$^+$ temperatures profiles appear smooth, both $^{12}$CO profiles show the presence of a dip at ${\approx}$120~au. This dip occurs exterior to the continuum edge and corresponds to a small vertical dip in molecular emission height at the same radius (Section \ref{sec:overview_emission_surfaces}). This temperature decrease could either be due to locally reduced gas surface density, in which the emission surface traces a deeper and thus cooler layer, or due to a change in heating or radiation properties related to the continuum edge. To better quantify these temperature dips, we use the same method as for the vertical substructures in Section \ref{sec:overview_emission_surfaces}. The dip in temperature for $^{12}$CO J=2--1 occurs at 115~au, while in $^{12}$CO J=3--2, it is at 127~au. Both features have comparable widths (${\approx}$20-30~au) and depths (${\lesssim}10\%$). A full listing of temperature substructure properties is provided in Appendix \ref{sec:appendix_Fnu_profiles}.

To better characterize the radial temperature profiles, we fitted each line with a power law profile as:

\begin{equation}
T = T_{100} \times \left(\frac{r}{\rm{100\,au}} \right)^{-q},
\end{equation}

\noindent where $q$ is the slope and T$_{100}$ is  the brightness temperature at 100~au. Table \ref{tab:radial_temperature_plaw_fits} shows the fitted parameters. Overall, the profiles have slopes of $q\approx0.4$-$0.8$, which is generally consistent with other disks \citep[e.g.,][]{Law22}, with $^{12}$CO J=3--2 having the steepest radial temperature profile.

\begin{deluxetable*}{lccccc}[p]
\tablecaption{Radial Temperature Profile Fits\label{tab:radial_temperature_plaw_fits}}
\tablewidth{0pt}
\tablehead{
\colhead{Line} & \colhead{r$_{\rm{fit, in}}$ (au)}  &\colhead{r$_{\rm{fit, out}}$ (au)} & \colhead{T$_{100}$ (K)} &  \colhead{q} &  \colhead{Feat.\tablenotemark{a}}} 
\startdata
$^{12}$CO J=2$-$1 & 72 & 204 & 32~$\pm$~0.3 & 0.57~$\pm$~0.03 & D115\\
$^{13}$CO J=2$-$1 & 75 & 164 & 18~$\pm$~0.5 & 0.44~$\pm$~0.10 & \\
$^{12}$CO J=3$-$2 & 70 & 216 & 31~$\pm$~0.2 & 0.79~$\pm$~0.02 & D127\\
HCO$^+$ J=4$-$3 & 46 & 161 & 22~$\pm$~0.2 & 0.43~$\pm$~0.02 & \\
\enddata
\tablenotetext{a}{Local temperature dips (D) labeled according to their approximate radial location in au. See Appendix \ref{sec:appendix_Fnu_profiles} for further details.}\end{deluxetable*}

\begin{figure*}[p]
\centering
\includegraphics[width=\linewidth]{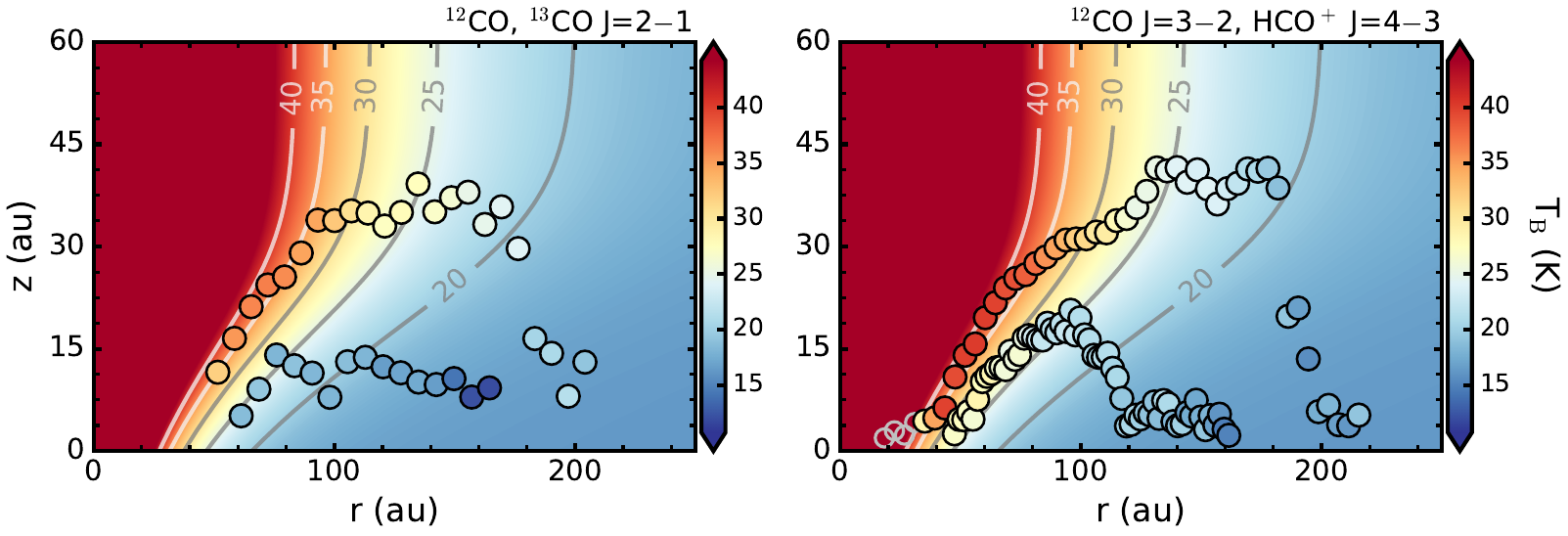}
\caption{Comparison of the measured temperatures (points) with the fitted 2D temperature structures (background) for the PDS~70 disk, as listed in Table \ref{tab:2D_temperature_fit_params}. The $^{12}$CO and $^{13}$CO J=2--1 (left) and $^{12}$CO J=3--2 and HCO$^+$ J=4--3 (right) are shown in separate panels for visual clarity but the 2D temperature was fit to both sets of lines simultaneously. The same color scale is used for the data and fitted model in each panel. Points excluded from the fits are show as hollow markers and contours show constant temperatures. The uncertainty of the temperature measurements, which is not shown here, can be found in Figure \ref{fig:2D_temp_surfaces}.}
\label{fig:2D_temp_fits}
\end{figure*}

\begin{deluxetable*}{lcccccccccccc}[p]
\tablecaption{Summary of 2D Temperature Structure Fits\label{tab:2D_temperature_fit_params}}
\tablewidth{0pt}
\tablehead{ \colhead{$T_{\rm{atm},0}$ (K)} & \colhead{$T_{\rm{mid}, 0}$ (K)} & \colhead{$q_{\rm{atm}}$}& \colhead{$q_{\rm{mid}}$} & \colhead{$z_0$ (au)} & \colhead{$\alpha$} & \colhead{$\beta$} }
\startdata
33$^{+0.2}_{-0.2}$ & 16$^{+0.3}_{-0.3}$ & $-$0.95$^{+0.02}_{-0.02}$ & $-$0.01$^{+0.04}_{-0.04}$ & 13$^{+0.2}_{-0.2}$ & 2.13$^{+0.03}_{-0.03}$ & 0.14$^{+0.03}_{-0.03}$\\
\enddata
\end{deluxetable*}

\subsubsection{2D Temperature Profiles} \label{sec:2D_temps}

Figure \ref{fig:2D_temp_surfaces} shows the thermal structure of the CO isotopologues and HCO$^+$ emitting layers as a function of ($r$, $z$). By combining all lines with derived surfaces, which trace different heights in the disk, we can construct a 2D model of the temperature distribution. Following \citet{Law21_MAPSIV}, we fit a two-layer temperature model in which disk midplane and atmosphere temperature are power laws smoothly connected with a hyperbolic tangent function \citep{Dartois03, Dullemond20}:

\begin{equation}
T_{\rm{mid}} (r) = T_{\rm{mid}, 0} \left( r / 100~\rm{au} \right)^{q_{\rm{mid}}}
\end{equation}

\begin{equation}
T_{\rm{atm}} (r) = T_{\rm{atm}, 0} \left( r / 100~\rm{au} \right)^{q_{\rm{atm}}}
\end{equation}

\begin{equation} \label{eqn:trig}
T^4 (r, z) =  T^4_{\rm{mid}} (r) + \frac{1}{2} \left[ 1 + \tanh \left( \frac{z - \alpha z_q(r)}{z_q(r)} \right) \right] T^4_{\rm{atm}} (r),
\end{equation}

\noindent where $z_q (r) = z_0 \left(r / 100~\rm{au} \right)^{\beta}$. The $\alpha$ parameter defines the height at which the transition in the tanh vertical temperature profile occurs, while $\beta$ describes how the transition height varies with radius.

We used the MCMC sampler in \texttt{emcee} \citep{Foreman_Mackey13} to estimate the following seven parameters: $T_{\rm atm,0}$, $q_{\rm atm}$, $T_{\rm mid,0}$, $q_{\rm mid}$, $\alpha$, $z_0$, and $\beta$. We used 256 walkers, which take 500 steps to burn in and then an additional 5000 steps to sample the posterior distribution function. Figure \ref{fig:2D_temp_fits} shows the fitted 2D temperature profiles and Table \ref{tab:2D_temperature_fit_params} lists the fitted parameter values and uncertainties, which are given as the 50th, 16th, and 84th percentiles from the marginalized posterior distributions, respectively.

\section{Discussion} \label{sec:discussion}

\subsection{Influence of Embedded Planets PDS~70b and 70c on Disk Vertical Structure} \label{sec:PDS_70bc}

The presence of embedded massive protoplanets may be expected to either locally or perhaps globally alter the vertical gas structure of protoplanetary disks, which in turn, would affect the properties of the material available to be accreted by those planets. The PDS~70 system provides an ideal environment to search for such effects. Here, we first discuss the potential presence of any local perturbations due to PDS~70b and 70c and then compare the overall vertical structure of the PDS~70 disk with other disks in the literature.

\subsubsection{Local Perturbations and Cavity Wall}

In the $^{12}$CO J=3--2 emission surface, we identify relatively flat ($z/r\approx0.1$) heights that overlap with the planet locations, i.e., within the central ${\approx}$45~au. This vertically-flat region extends to the edge of the cavity wall, where the heights sharply rise again to $z/r\approx 0.3$ (Figure \ref{fig:figure_gallery_r_v_z_12CO}). This is the same region that shows a deep drop in brightness temperature, reaching as low as ${\approx}$15~K (Figure \ref{fig:figure_temp}). 

The flat emitting heights observed within ${\approx}$45~au are consistent with expectations of a transition disk with a deep gas- and dust-depleted central cavity, which in the case of the PDS~70 disk has been carved out by PDS~70b and 70c \citep[][]{Bae19, Portilla_inprep}. Line emission surfaces are tracing regions where optical depths reach unity. Thus, in this region near the planets, the overall line optical depth decreases due to the reduced gas surface density and the surface is pushed deeper into the disk, i.e., close to the midplane, which naturally explains both the dip in vertical heights and the lower brightness temperatures. This is the first transition disk where this effect is directly observed in line emitting heights (see Section \ref{sec:discussion_LkCa15} for more details). For all other lines besides $^{12}$CO J=3--2, no estimates of vertical heights within the central cavity were possible due to insufficient angular resolution or intrinsically flat emitting heights likely due to the low gas surface densities.

\subsubsection{Outer Disk Vertical Structure of PDS 70 in Context}

Although we do not see any clear local perturbations around PDS~70b or 70c beyond that of the cavity wall, the overall structure of the gas disk may still be influenced by the presence of both planets in the form of, e.g., dynamical or planet-disk interactions. In particular, here we focus on the outer disk vertical gas structure. One way to assess this is to compare the PDS~70 disk with others in the literature with similar constraints on their line emitting heights.

\begin{figure*}[t]
\centering
\includegraphics[width=0.9\linewidth]{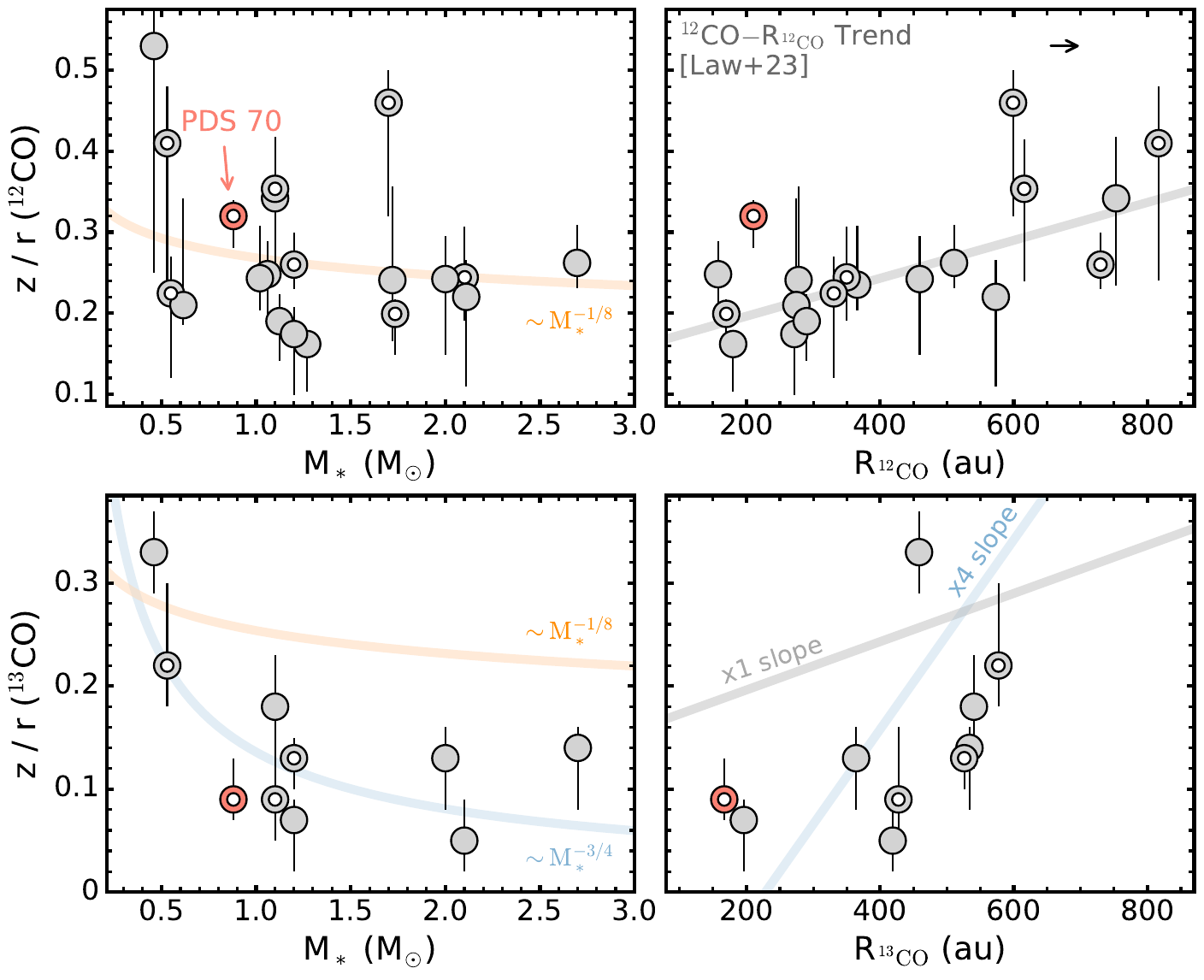}
\caption{Comparison of the characteristic $^{12}$CO (top row) and $^{13}$CO (bottom row) emission heights of the PDS~70 disk (in pink) vs other disks in the literature \citep[][and see references therein]{Law23}. Transition disks are shown as symbols with a hollow center. All stellar masses are dynamical, $z/r$ values are extracted homogeneously via directly-extracted emission surfaces, and disk sizes represent the radius enclosing 90\% of the total flux. The dynamical stellar mass of PDS~70 is taken from \citet{Keppler19}. Several scaling relations, i.e., line emission heights scale with gas pressure scale height, and a previously-identified positive $^{12}$CO height trend and disk size are labeled \citep[see][for more details]{Law22, Law23}.}
\label{fig:ZR_Gallery}
\end{figure*}

To do this, we computed the characteristic $z/r$ heights of each line in the PDS~70 disk following the procedure outlined in \citet{Law22}, i.e., mean $z/r$ within the steeply-rising portion of the surfaces, for consistent comparison with literature values. Table \ref{tab:emission_surf} lists the computed values. We first consider the CO isotopologues and then HCO$^+$: \\

\noindent \textit{CO Isotopologues}: Figure \ref{fig:ZR_Gallery} shows the characteristic $z/r$ emission heights of $^{12}$CO and $^{13}$CO versus the stellar mass and disk size compiled from consistently derived heights \citep{Law23}. We also include several expected scaling relations, i.e., assuming line emission heights scale with gas pressure scale heights, and a previously-identified positive $^{12}$CO-R$_{^{12}\rm{CO}}$ trend \citep[for more details, see][]{Law22,Law23}. With respect to the literature sample, the PDS~70 disk appears quite typical in terms of its $^{12}$CO and $^{13}$CO emitting heights compared to disks with similar host stellar masses and disk sizes. This is true for all disks, or if we only consider the sub-sample of transition disks. Thus, the two planets in the PDS~70 disk do not seem to be substantially altering the heights of the line emission surfaces via, e.g., planet-disk interactions. However, we note the caveat that many of the literature disks used as the comparison sample also have indirect evidence for the presence of embedded planets. Thus, the comparison of emitting heights in the PDS~70 disk versus those disks without planets is not necessarily as straightforward as Figure \ref{fig:ZR_Gallery} suggests, but such a comparison nonetheless establishes that the PDS~70 disk has a vertical gas structure that is typical of the large, resolved disks thus far studied in detail with ALMA. 

The presence of embedded planets could, for instance, have resulted in flatter emitting surfaces due to the removal or redistribution of gas via vertical flows \citep[e.g.,][]{Teague19Natur, Yu21, Galloway23}, or alternatively, driven dynamical interactions which would inflate the disk gas vertical distribution \citep[e.g.,][]{Montesinos21, Kuo22}. However, the close-in radial locations of the PDS~70 planets may provide an explanation, in which the vertical disk gas distribution is decoupled from planet formation occurring in the inner few 10s of au. 

\begin{figure*}[t]
\centering
\includegraphics[width=\linewidth]{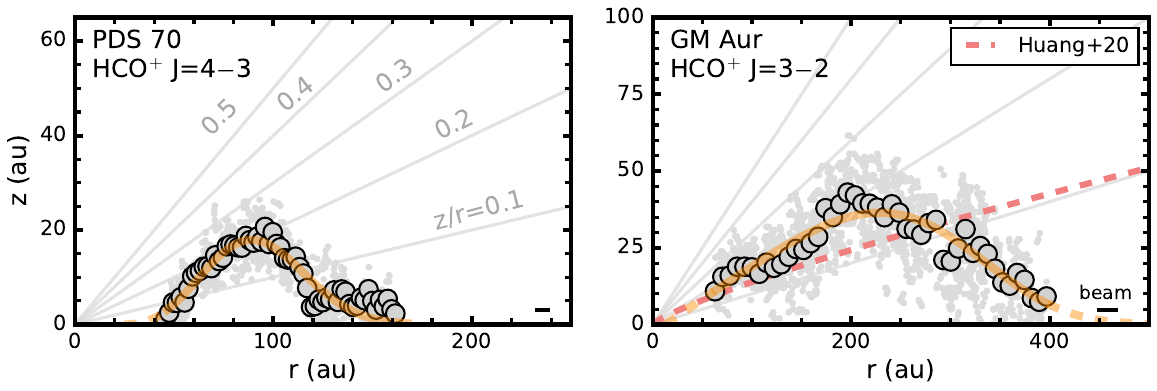}
\includegraphics[width=\linewidth]{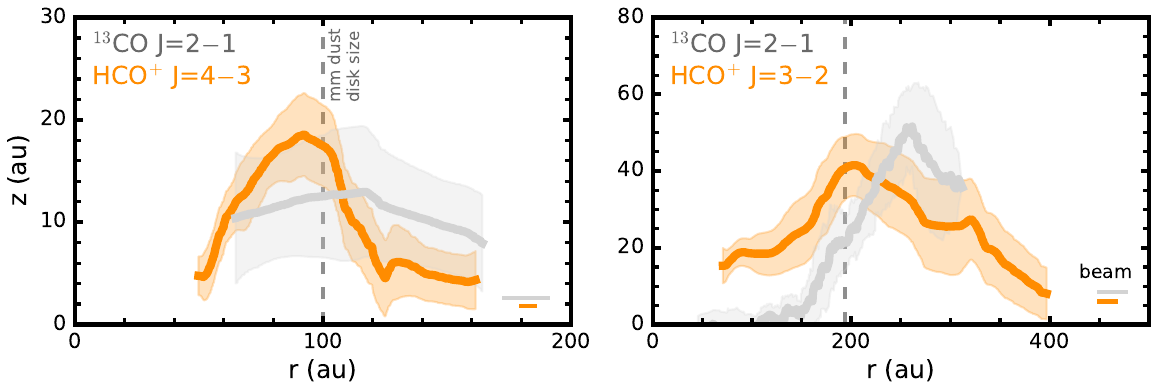}
\caption{Top: Comparison of HCO$^+$ surfaces in the PDS~70 (left) and GM~Aur (right) disks. The GM~Aur data were obtained from \citet{Huang20} and we extracted the emission surface as in Section \ref{sec:methods_sub_surfextr}. The single power law fit from \citet{Huang20} is also shown for reference. Otherwise, as in Figure \ref{fig:figure_gallery_r_v_z_12CO}. Bottom: Zoom-in on the HCO$^+$ (orange) and $^{13}$CO (gray) emission surfaces. The mm continuum disk size is marked by a vertical line. The $^{13}$CO surface in the GM~Aur disk is from \citet{Law21_MAPSIV}.}
\label{fig:HCOp_GMAur}
\end{figure*}

Several previous works \citep{Paneque23, Izquierdo23} have identified vertical variations in line emitting heights that are coherent across several tracers at the radial locations of suspected planetary companions or kinematic deviations. In the PDS~70 disk, we find vertical variations at radii between ${\approx}$95-120~au in both $^{12}$CO and $^{13}$CO. However, no evidence for kinematic deviations or additional embedded planets exists in the outer disk of PDS~70. Thus, it is more likely that these vertical substructures are instead related to the edge of the mm continuum disk due to, e.g., changing radiation fields at the dust edge.  \\

\noindent \textit{HCO$^+$}: Only a few measurements of HCO$^+$ emitting heights exist in protoplanetary disks. \citet{Paneque23} derived heights of the J=1--0 line in the MAPS disks \citep{Oberg21_MAPSI} and for the J=4--3 line in the WaOph~6 disk, while \citet{Huang20} constrained the J=3--2 emitting surface in the GM~Aur disk. The J=1--0 lines showed typical heights of $z/r \lesssim 0.1$, but cannot be directly compared to PDS~70, because the difference in excitation properties between the J=4--3 ($\rm{E}_{\rm{u}}\approx$~43~K) and J=1--0 ($\rm{E}_{\rm{u}}\approx$~4~K) lines mean that we do not necessarily expect both lines to trace the same disk vertical layers. The WaOph~6 measurement, although using the same transition as in PDS~70, suffered from high uncertainties with a possible $z/r$ ranging from 0.1 to 0.5 due to the coarse beam size (${\approx}$0\farcs3) and lower SNR of the observations. This range of heights is consistent with what we measure in PDS~70 but does not permit a detailed comparison of the emitting surfaces. The most similar previous measurement is in the transition disk GM~Aur, where \citet{Huang20} used high angular resolution observations to fit a single power law profile to the HCO$^+$ J=3--2 emission surface and found an approximate height of $z/r\gtrsim0.1$. Although not identical, the J=3--2 (E$_{\rm{u}}\approx$26~K) line has more comparable excitation properties to that of the J=4--3 line. Moreover, both PDS~70 and GM~Aur are T~Tauri stars of similar spectral types and host transition disks with central cavities of comparable size with the main difference being that GM~Aur is at least twice as large in overall disk size in both its mm dust and $^{12}$CO line emission. For consistency, we re-fit the HCO$^+$ J=3--2 data in the GM~Aur disk from \citet{Huang20} using an exponentially-tapered power-law with the same procedure as in Section \ref{sec:methods_sub_surfextr}.

Figure \ref{fig:HCOp_GMAur} shows the resultant emission surface, with best-fit parameters of $z_0=0\farcs22^{+0.02}_{-0.02}$, $\phi=0.99_{-0.17}^{+0.18}$, $r_{\rm{taper}}=1\farcs93_{-0.10}^{+0.09}$, $\psi=4.01_{-0.73}^{+0.61}$ and $r_{\rm{cavity}}=0\farcs08_{-0.06}^{+0.08}$. Our emission surface is higher than the one derived in \citet{Huang20} at all radii within 250~au, which is due to our surface extraction method being able to more accurately track individual pixels in the channels.

HCO$^+$ is emitting from similar disk vertical regions ($z/r \approx 0.2$) in both the GM~Aur and PDS~70 disks, and thus, similar to the CO isotopologues, PDS~70 does not appear to show any significant differences, which could be ascribed to the influence of PDS~70b or 70c. As noted in Section \ref{sec:overview_emission_surfaces} and shown in Figure \ref{fig:HCOp_GMAur}, the relative heights of the $^{13}$CO J=2--1 and HCO$^+$ J=4--3 surfaces cross over beyond ${\sim}$100~au. We find the same trend in the GM~Aur disk, with the HCO$^+$ surface located at higher altitudes than $^{13}$CO in the inner ${\sim}$200~au, while at large radii, the $^{13}$CO is coming from a higher disk layer than HCO$^+$. The cross-over point of these surfaces occurs directly exterior to bulk of the mm continuum emission in both disks, which suggests that this effect may be due to a change in illumination at the continuum edge. When combined with the lower densities expected at larger disk radii, this, in turn, would likely alter how far ionizing photons can penetrate in the disk and could thus naturally explain the lower heights of the HCO$^+$ surfaces in the outer disk. Although intriguing, with only two sources, it is difficult to make any additional conclusions about the vertical distribution of HCO$^+$ in protoplanetary disks more generally. We stress that high-angular resolution HCO$^+$ observations in additional disks, especially those with different properties than PDS~70 and GM~Aur, are urgently needed to set stringent constraints on disk ionization structure.

\subsection{Comparison to the LkCa~15 Transition Disk} \label{sec:discussion_LkCa15}

Here, we compare the PDS~70 disk to that of LkCa~15, which is one of the few transition disks whose vertical emission structure has been mapped in detail with multiple molecular lines \citep{Leemker22, Law23}. LkCa~15 has a similar stellar spectral type (K5) and age ($\sim$5~Myr) \citep{Donati19} and while the LkCa~15 disk is much larger than that of PDS~70 in both mm dust continuum and CO isotopologue emission by a factor of ${\approx}$2-3, both disks have nearly identical central cavity sizes \citep[e.g.,][]{Pietu06, Facchini20}. Moreover, both disks have comparable gas temperatures within 200~au and similar $^{12}$CO and $^{13}$CO emission layer $z/r$ heights \citep{Leemker22, Law23}. Thus, they are ideal systems to compare how line emitting heights change across their central cavities.

\begin{figure*}[!htpb]
\centering
\includegraphics[width=0.85\linewidth]{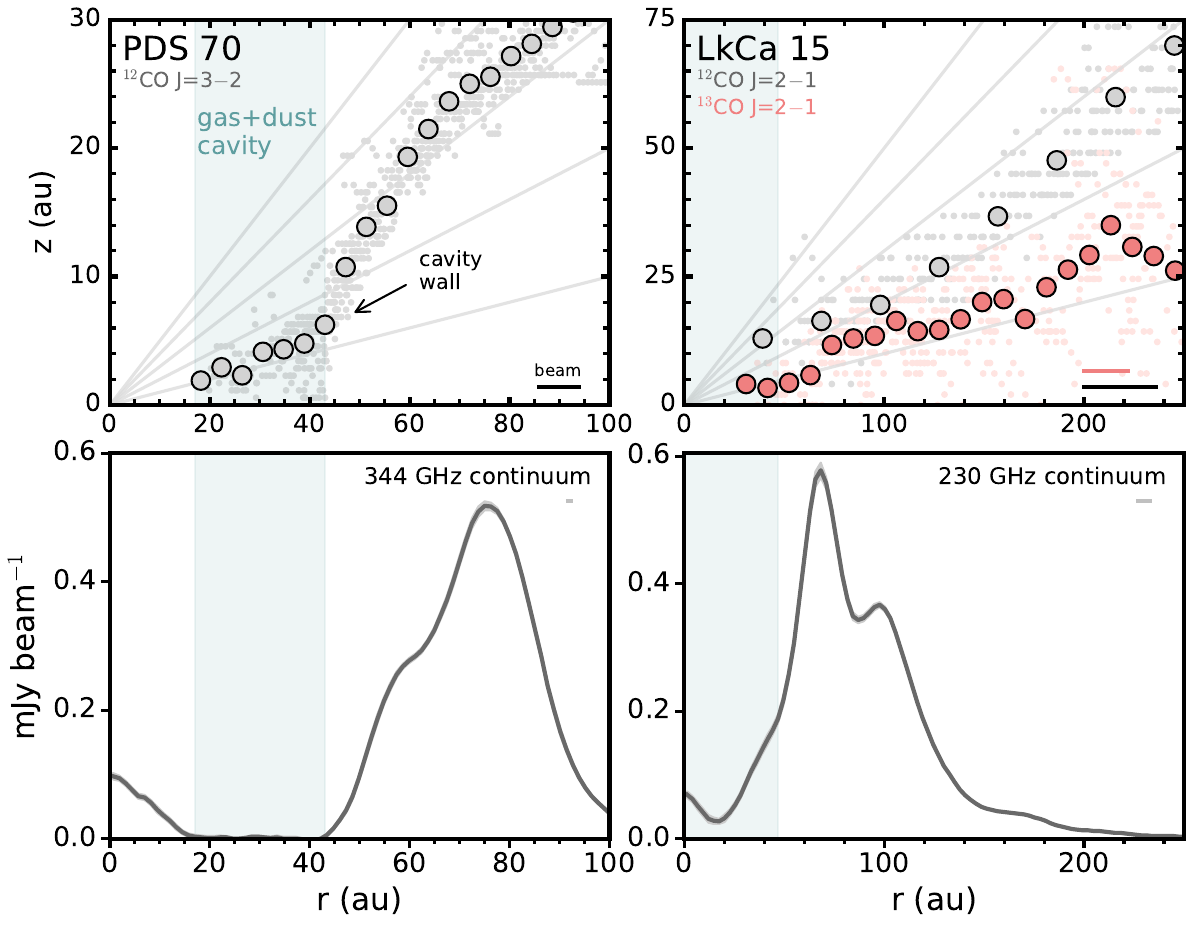}
\caption{Zoomed-in emission surfaces of transition disks around PDS~70 and LkCa~15 (\textit{top row}) compared to azimuthally-averaged millimeter continuum radial profiles (\textit{bottom row}). Lines of constant $z/r$ from 0.1 to 0.5 are shown in gray. Emission surfaces and continuum profiles in LkCa~15 are from \citet{Leemker22, Law23} and \citet{Facchini20}, respectively. Only the PDS~70 disk shows an increase in line emitting heights at the location of the cavity wall.}
\label{fig:compare_w_LkCa15}
\end{figure*}

Figure \ref{fig:compare_w_LkCa15} shows line emission surfaces in each disk zoomed into the location of the central cavity, as indicated in the corresponding azimuthally-averaged mm continuum radial profile. The PDS~70 disk shows an increase in emitting heights in $^{12}$CO J=3--2 associated with the edge of the cavity wall, while no such effect is seen in LkCa~15. While in Figure \ref{fig:compare_w_LkCa15}, we show the J=2--1 lines of $^{12}$CO and $^{13}$CO for the LkCa~15 disk, emitting height measurements also exist for $^{12}$CO J=3--2 and $^{13}$CO J=6--5 \citep{Leemker22}, but neither of these surfaces show a similar height increase as in PDS~70.

However, the absence of this cavity wall effect in the LkCa~15 disk is likely an observational limitation, rather than indicating any fundamental differences in inner disk structure. For instance, in Figure \ref{fig:compare_w_LkCa15}, the $^{12}$CO J=2--1 beam is ${\approx}$0\farcs3, which at the distance of LkCa~15 \citep[157~pc;][]{Gaia21}, corresponds to a physical scale of ${\approx}$50~au, which is approximately the size of the central cavity. In comparison, our $^{12}$CO J=3--2 observations of PDS~70 trace physical scales of ${\approx}$11-15~au, a factor of several times better. We note that while the $^{13}$CO J=6--5 surface in LkCa~15 from \citet{Leemker22} was derived from observations of comparable angular resolution, emitting heights interior to ${\approx}$40~au could not be extracted. Moreover, $^{13}$CO originates at lower elevations than that of $^{12}$CO, which means that this height increase at the cavity wall would be intrinsically more difficult to observe using $^{13}$CO.

Emission from $^{12}$CO J=2--1, 3--2 and $^{13}$CO J=6--5 have been detected in the central cavity of LkCa~15 \citep{Jin19, Leemker22, Law23}, which suggests that emission height constraints here would be possible with sufficiently sensitive and high-angular-resolution data. Additionally, \citet{Leemker22} found that the gas column density in the dust cavity (${\approx}$45~au) of LkCa~15 drops by a factor of ${>}$2 compared to the outer disk, with an additional order-of-magnitude decrease inside 10~au. In comparison, the PDS~70 cavity shows a higher gas depletion (${\gtrsim}$10$\times$) over its full cavity \citep{Portilla_inprep}. Thus, while we still expect flatter emission surfaces in the LkCa~15 cavity, this cavity wall feature may be less pronounced in the LkCa~15 disk due to its more modest gas density decrease. Nonetheless, observations from the ongoing exoALMA Large Program will probe similar physical resolutions as in the PDS~70 disk in several transition disks, including LkCa~15, and should be able to confirm if such changes in emitting heights at cavity walls are commonplace or if the PDS~70 disk structure is unique.

\subsection{CO Snowline Location} \label{sec:13CO_vs_params}

The elemental carbon-to-oxygen (C/O) ratio is of vital importance to understand the accretion history of protoplanets and connect disk chemistry with planetary organic compositions and atmospheres \citep[e.g.,][]{Madhusudhan19, Oberg21PhR}. While chemical evolution may alter the C/O ratio \citep[e.g.,][]{Eistrup18, Cridland19, Krijt20}, the condensation of major volatile species across the disk has the most dramatic influence on the elemental ratio of gas-phase material \citep[e.g.,][]{Oberg11}. These condensation fronts, often referred to as snowlines, occur at specific temperatures and thus, with detailed knowledge of the disk temperature structure, it is possible to estimate the location of these snowlines directly. Here, we focus on the CO snowline, which is readily observationally-accessible at spatial scales probed by ALMA.

\begin{figure}[t]
\centering
\includegraphics[width=\linewidth]{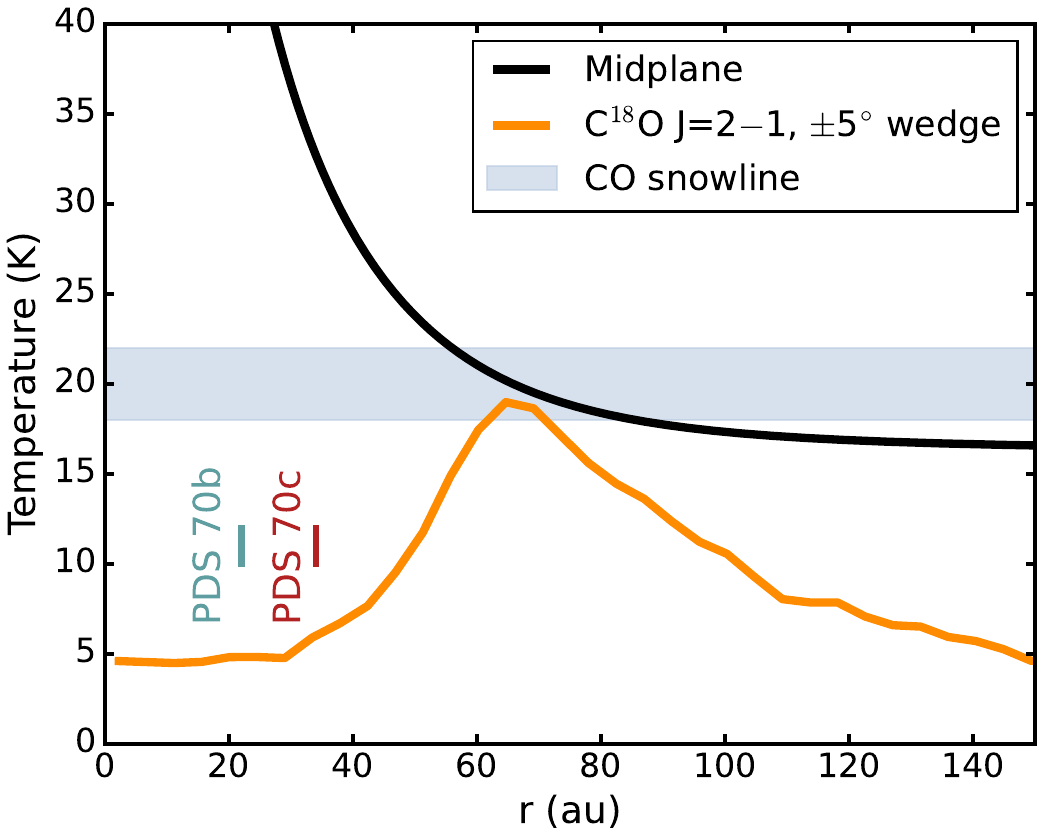}
\caption{Midplane gas temperature as a function of radius in the PDS~70 disk as inferred from the 2D empirical temperature fit. The blue horizontal region indicates a range of CO freezeout temperatures (18-22~K), which corresponds to a midplane snowline radius of ${\approx}$56-85~au. The peak brightness temperature profile of C$^{18}$O J=2--1 extracted in a narrow $\pm$5$^{\circ}$ wedge along the disk major axis is shown in orange.} 
\label{fig:CO_snowline}
\end{figure}

Figure \ref{fig:CO_snowline} shows the midplane gas temperature as inferred from the 2D empirical fit (Figure \ref{fig:2D_temp_fits}). It is important to note that the midplane temperature from this fit is an extrapolation from higher disk heights since we lack lines that directly trace the disk midplane interior to a radius of 100~au. As a result, we also show the C$^{18}$O J=2--1 peak brightness profile temperature in Figure \ref{fig:CO_snowline}, which we expect to be emitting at, or near, the disk midplane (Section \ref{sec:overview_emission_surfaces}). Beyond the central cavity, the C$^{18}$O brightness temperature is generally consistent with our estimated midplane temperature and is ${\approx}$5-10~K colder at larger radii (${\gtrsim}$100~au). As C$^{18}$O is not expected to be fully optically thick, especially at large radii, it only provides a lower limit on the gas temperature, which is consistent with our warmer midplane temperature estimates.

If we adopt a reasonable range of possible CO freeze-out temperatures from 18-22~K \citep[e.g.,][]{Collings04, Oberg05, Rafa14, Facchini17}, we estimate an approximate snowline radius of ${\approx}$56-85~au. Despite this range, we can confidently conclude that both PDS~70b and 70c are located interior the CO snowline. Thus, the nearby gas from which both planets are accreting likely has a super-stellar\footnote{Based on its photospheric abundances of carbon and oxygen, PDS~70 has a C/O ratio of 0.44 $\pm$ 0.19 \citep{Suarez18,Cridland23}.} C/O ratio. Depending on the exact location of the CO$_2$ snowline, this gas likely has a C/O of ${\lesssim}$1.0 \citep{Oberg11}, with the exact value depending on the details of radial transport and chemical processing \citep[e.g.,][]{Krijt18, Krijt20}. From our midplane temperature estimate, we would estimate a CO$_2$ snowline location of ${\approx}$21-26~au (assuming freeze-out temperatures of 42-52~K), which is consistent with C/O $\approx 1$, with the caveat that these are extrapolations to radii within the inner cavity where we lack direct temperature measurements.

Overall, this is consistent with the findings of \citet{Facchini21PDS}, who found that the PDS~70 molecular layer hosts a high C/O ratio inferred from bright C$_2$H line fluxes and their lower limit on the CS/SO column density ratio \citep[e.g.,][]{Bergin16, Semenov18, Miotello19, Legal21}. The fact that both planets in the PDS~70 disk are likely accreting gas with a high C/O ratio provides important context for interpreting future atmospheric planetary measurements from, e.g., JWST, and for properly modeling their formation histories within the PDS~70 disk \citep[e.g.,][]{Cridland23}.

Further observations of temperature tracers, such as additional J lines of CO isotopologues \citep[e.g.,][]{Leemker22}, especially those closer to the midplane and in the inner disk would allow for a better constrained temperature model and lead to a more accurate prediction of the snowline location. Complementary to this, observations of N$_2$H$^+$, which has been shown to be an accurate observational tracer of the CO snowline \citep{Qi19}, would provide an independent estimate of the CO snowline location. If combined with our temperature-based estimate, this would provide a unique opportunity to measure the CO binding energy in an astrophysical context and provide important and independent context to laboratory-based estimates.

\section{Conclusions} \label{sec:conlcusions}

We present observations of a suite of CO isotopologue and HCO$^+$ lines toward the PDS~70 disk at high angular resolution (${\approx}0\farcs1$). We extracted line emission surfaces and conclude the following:

\begin{enumerate}
    \item $^{12}$CO emission surfaces originate in elevated disk regions ($z/r\approx0.3$), while the less abundant $^{13}$CO emitting heights trace deeper disk layers ($z/r\approx0.1$). We also derived emitting heights for the HCO$^{+}$ J=4--3 line, which arises from $z/r \approx 0.2$ in the inner 100~au and shows an extended, flat ($z/r\lesssim0.1$) component at larger radii. 
    \item In the $^{12}$CO J=3--2 line, we clearly resolve the vertical dip and steep rise in emitting heights at ${\approx}$15-45~au due to the cavity wall in the PDS~70 disk. This is the first transition disk where this effect has been directly seen in emitting heights.
    \item The emitting heights of the CO isotopologues in the outer disk of PDS~70 appear typical for its stellar mass and disk size and do not appear to be substantially altered by the presence of its two embedded planets. Overall, this suggests that the outer disk is largely decoupled from the planet formation occurring in the inner few 10s of au.
    \item By combining CO isotopologue and HCO$^+$ lines, we derive an empirical 2D temperature structure for the PDS~70 disk. Using this, we estimate an approximate CO midplane snowline radius of ${\approx}$56-85~au, which implies that both PDS~70b and 70c are located well interior to the CO snowline and suggests they are both accreting from gas with a C/O ratio of ${\approx}$1.0.
\end{enumerate}

The authors thank the anonymous referee for valuable comments that improved both the content and presentation of this work. This paper makes use of the following ALMA data: ADS/JAO.ALMA\#2015.1.00888.S, 2017.A.00006.S, 2017.1.01151.S, 2018.A.00030.S, 2018.1.01230.S, and 2019.1.01619.S. ALMA is a partnership of ESO (representing its member states), NSF (USA) and NINS (Japan), together with NRC (Canada), MOST and ASIAA (Taiwan), and KASI (Republic of Korea), in cooperation with the Republic of Chile. The Joint ALMA Observatory is operated by ESO, AUI/NRAO and NAOJ. The National Radio Astronomy Observatory is a facility of the National Science Foundation operated under cooperative agreement by Associated Universities, Inc.

Support for C.J.L. was provided by NASA through the NASA Hubble Fellowship grant No. HST-HF2-51535.001-A awarded by the Space Telescope Science Institute, which is operated by the Association of Universities for Research in Astronomy, Inc., for NASA, under contract NAS5-26555. S.F. is funded by the European Union under the European Union's Horizon Europe Research \& Innovation Programme 101076613 (UNVEIL). Views and opinions expressed are however those of the author(s) only and do not necessarily reflect those of the European Union or the European Research Council. Neither the European Union nor the granting authority can be held responsible for them. This project has received funding from the European Research Council (ERC) under the European Union's Horizon 2020 research and innovation programme (PROTOPLANETS, grant agreement No.~101002188).

\facilities{ALMA}

\software{Astropy \citep{astropy_2013,astropy_2018,Astropy_22}, \texttt{bettermoments} \citep{Teague18_bettermoments}, CASA \citep{McMullin_etal_2007, CASA}, \texttt{cmasher} \citep{vanderVelden20}, \texttt{disksurf} \citep{disksurf_Teague}, \texttt{GoFish} \citep{Teague19JOSS}, \texttt{keplerian\_mask} \citep{rich_teague_2020_4321137}, Matplotlib \citep{Hunter07}, NumPy \citep{vanderWalt_etal_2011}, scikit-image \citep{vanderWalt14_scikit}, SciPy \citep{Virtanen_etal_2020}}



\appendix

\section{Line Emission Channel Maps} \label{sec:appendix_channel_maps}

A complete gallery of channel maps for all CO isotopologue lines and HCO$^+$ J=4--3 is shown in Figure Set 1, which is available in the electronic edition of the journal.

\begin{figure*}[]
\centering
\includegraphics[width=\linewidth]{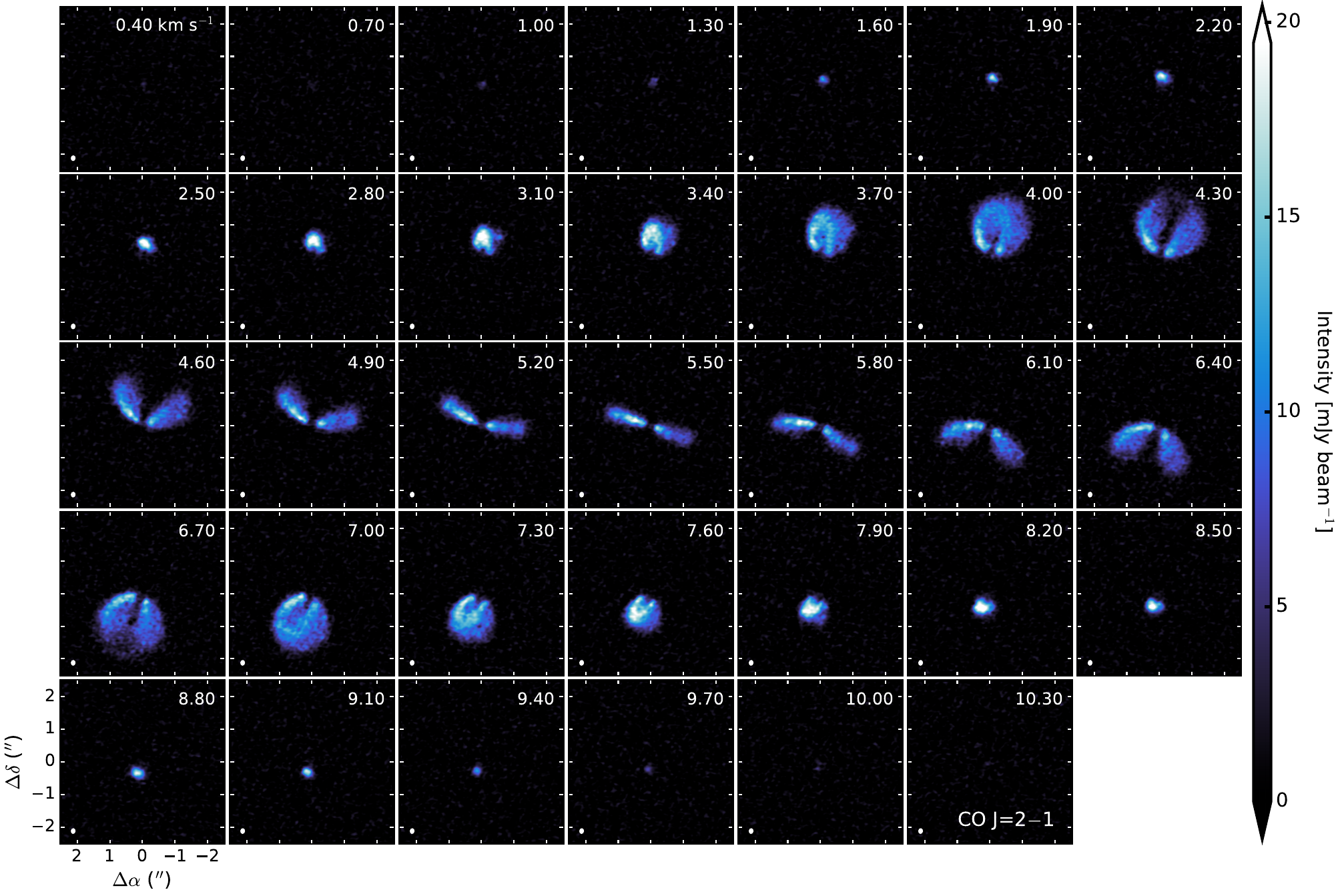}
\caption{Channel maps of the $^{12}$CO J=2--1 emission of the PDS~70 disk. The synthesized beam is shown in the lower left corner of each panel and the LSRK velocity in km~s$^{-1}$ is printed in the upper right.}
\label{fig:Figure13_1}
\end{figure*}

\section{Radial Temperatures vs Peak Brightness Temperatures} \label{sec:appendix_Fnu_profiles}

\begin{figure*}[]
\centering
\includegraphics[width=\linewidth]{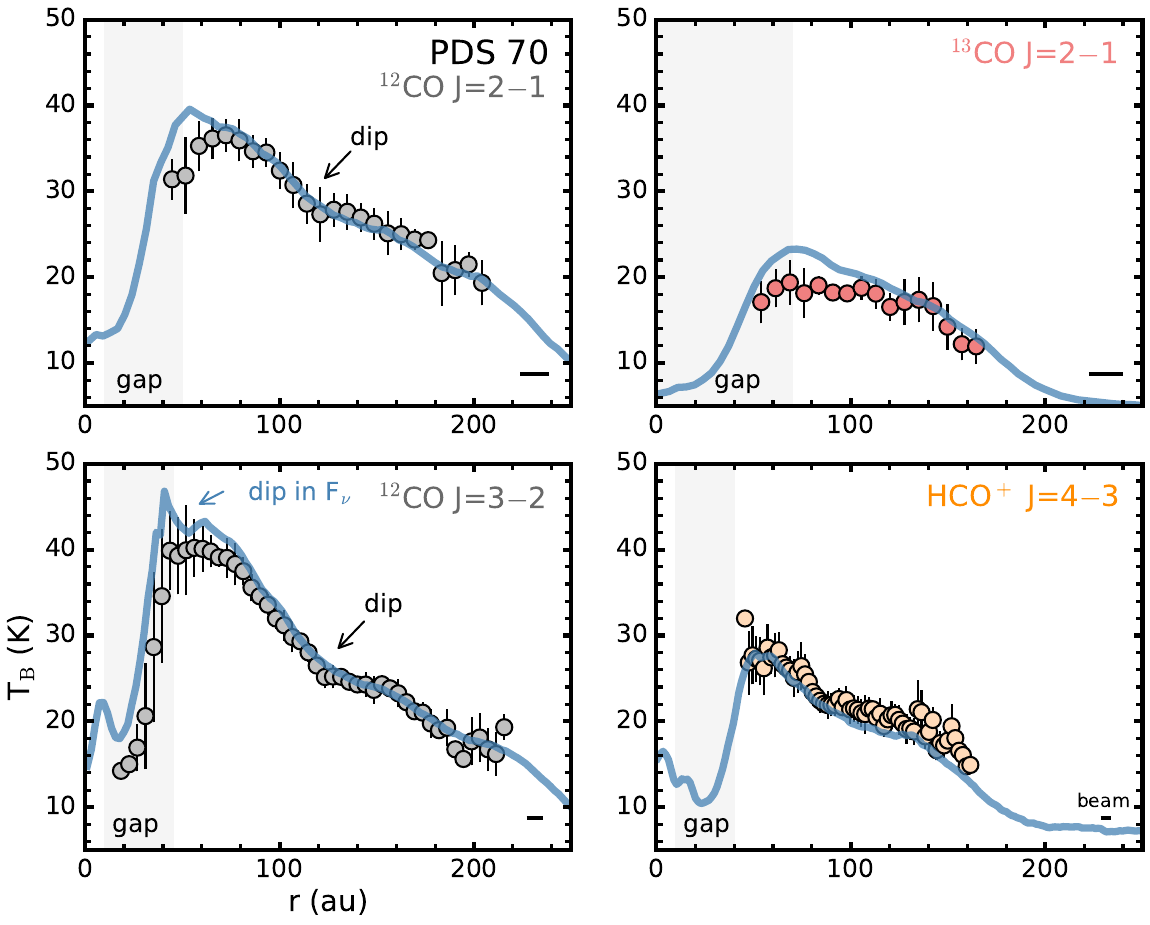}
\caption{Peak brightness (blue curve) vs radial temperature profiles (points) computed along the emission surfaces. The azimuthally-averaged peak intensity  profiles are generated by deprojecting along the derived emission surfaces.}
\label{fig:app:Fnu}
\end{figure*}

We extracted radial profiles from the peak intensity maps (Figure \ref{fig:overlaid_Fnu}) by adopting the inferred emission surfaces and following the same procedure as in Section \ref{sec:vertical_substr_vs_mm_cont}. In Figure \ref{fig:app:Fnu}, we compare them to the radial temperatures inferred directly from the emission surfaces. Overall, we find excellent agreement between the two curves. The most notable differences occur in the inner 100~au of the $^{13}$CO J=2--1 profile, where the peak intensities have temperatures which are at most ${\lesssim}$5~K larger than the surface-extracted ones. We also note that the dips identified at ${\approx}$120~au in both $^{12}$CO J=2--1 and J=3--2 from Figure \ref{fig:figure_temp} are also present in the peak intensities.

We also find a small dip in the peak intensity profiles of $^{12}$CO J=3--2 at ${\sim}50$~au that is not apparent in the radial brightness temperature profiles. This dip was identified previously in integrated intensity by \citet{Keppler19} and was attributed to continuum absorption of the back side of the disk, since it appears coincident with the mm continuum gap. However, given its appearance in the peak intensity profile, which should be less affected by continuum absorption, this suggests that it is a real feature and may hint at a true gas and dust gap. While we do not see this feature in any of the other CO lines, this is perhaps not surprising given the small width of this feature and the larger beam sizes of the other CO lines. Additional observations at high angular resolution of other CO isotopologues are necessary to confirm the nature of this feature.

As noted in Section \ref{sec:radial_temperatures}, we also determined the properties of the temperature dips seen in Figures \ref{fig:figure_temp} and \ref{fig:app:Fnu} using the same method as for the vertical substructures, i.e., via removing a local baseline \citep{Law21_MAPSIV}. Table \ref{tab:temperature_substructures} shows the properties of each of these features, including the 50~au dip in $^{12}$CO J=3--2 identified in the peak intensity radial profiles.

\begin{deluxetable*}{llccccccc}[]
\tablecaption{Properties of Radial Temperature Dips \label{tab:temperature_substructures}}
\tablewidth{0pt}
\tablehead{\colhead{Line} & \colhead{Feature} & \colhead{$r_0$ (au)} & \colhead{Width (au)} & \colhead{$\Delta$T (K)} & \colhead{Depth}}
\startdata
$^{12}$CO J=2$-$1 & D115 & 115 $\pm$ 0.3 & 22 $\pm$ 11 &  2.8 $\pm$ 0.9 & 0.09 $\pm$ 0.003\\
$^{12}$CO J=3$-$2 & D50\tablenotemark{a} & 50 $\pm$ 2 & 13 $\pm$ 0.0 &  2.3 $\pm$ 0.3 & 0.07 $\pm$ 0.002\\
$^{12}$CO J=3$-$2 & D127 & 127 $\pm$ 1 & 33 $\pm$ 12 &  2.1 $\pm$ 0.7 & 0.08 $\pm$ 0.01\\
\enddata
\tablenotetext{a}{Computed from the peak intensity radial profile.}
\end{deluxetable*}

\newpage
\clearpage


\bibliography{PDS70_surfaces}{}
\bibliographystyle{aasjournal}



\end{document}